\documentclass[acmsmall]{acmart}
\usepackage{booktabs,makecell,multirow}
\usepackage{subfig}
\usepackage{pifont}
\usepackage{wrapfig}
\usepackage{adjustbox}
\usepackage{tablefootnote}

\usepackage{atbegshi}
\AtBeginDocument{\AtBeginShipoutNext{\AtBeginShipoutDiscard}}

\setcopyright{acmlicensed}
\copyrightyear{2024}
\acmYear{2024}
\acmDOI{XXXXXXX.XXXXXXX}

\acmVolume{37}
\acmNumber{4}
\acmArticle{111}
\acmMonth{12}

\begin{document}

\title{A Survey on Error-bounded Lossy Compression for Scientific Datasets}

\author{Sheng Di}
\email{sdi1@anl.gov}
\orcid{0000-0002-7339-5256}
\affiliation{%
  \institution{Argonne National Laboratory}
  \streetaddress{9700 Cass Ave.}
  \city{Lemont}
  \state{Illinois}
  \country{USA}
  \postcode{60439}
}

\author{Jinyang liu}
\email{jliu447@ucr.edu}
\affiliation{%
  \institution{University of California, Riverside}
  \city{Riverside}
  \state{California}
  \country{USA}
  \postcode{27696}
}

\author{Kai Zhao}
\email{kai.zhao@fsu.edu}
\affiliation{%
  \institution{Florida State University}
  \streetaddress{600 W. College Ave.}
  \city{Tallahassee}
  \country{USA}
  \postcode{32306}
}

\author{Xin Liang}
\email{xliang@uky.edu}
\affiliation{%
 \institution{University of Kentucky}
 \streetaddress{410 Administration Dr.}
 \city{Lexington}
 \state{Kentucky}
 \country{USA}
 \postcode{40506}
 }

\author{Robert Underwood}
\email{runderwood@anl.gov}
\affiliation{%
  \institution{Argonne National Laboratory}
  \streetaddress{9700 Cass Ave.}
  \city{Lemont}
  \state{Illinois}
  \country{USA}
  \postcode{60439}
}

\author{Zhaorui Zhang}
\email{zhaorui.zhang@polyu.edu.hk}
\affiliation{%
  \institution{The Hong Kong Polytechnic University}
  \streetaddress{11 Yuk Choi Rd}
  \city{Hong Kong}
  \country{China}
}

\author{Milan Shah}
\email{mkshah5@ncsu.edu}
\affiliation{%
  \institution{North Carolina State University}
  \streetaddress{2610 Cates Ave.}
  \city{Raleigh}
  \state{North Carolina}
  \country{USA}
  \postcode{27695}
}

\author{Yafan Huang}
\email{yafan-huang@uiowa.edu}
\orcid{0000-0001-7370-6766}
\affiliation{%
  \institution{University of Iowa}
  \streetaddress{201 S. Clinton St.}
  \city{Iowa City}
  \state{Iowa}
  \country{USA}
  \postcode{52246}
}

\author{Jiajun Huang}
\email{jhuan380@ucr.edu}
\orcid{0000-0001-5092-3987}
\affiliation{%
  \institution{University of California, Riverside}
  \streetaddress{900 University Ave.}
  \city{Riverside}
  \state{California}
  \country{USA}
  \postcode{92521}
}

\author{Xiaodong Yu}
\email{xyu38@stevens.edu}
\affiliation{%
  \institution{Stevens Institute of Technology}
  \streetaddress{1 Castle Point Terrace}
  \city{Hoboken}
  \state{New Jersey}
  \country{USA}
  \postcode{07030}
}

\author{Congrong Ren}
\email{ren.452@osu.edu}
\affiliation{%
  \institution{The Ohio State University}
  \streetaddress{2015 Neil Ave.}
  \city{Columbus}
  \state{Ohio}
  \country{USA}
  \postcode{43210}
}

\author{Hanqi Guo}
\email{guo.2154@osu.edu}
\affiliation{%
  \institution{The Ohio State University}
  \streetaddress{2015 Neil Ave.}
  \city{Columbus}
  \state{Ohio}
  \country{USA}
  \postcode{43210}
}

\author{Grant Wilkins}
\email{gfw27@cam.ac.uk}
\affiliation{%
  \institution{University of Cambridge}
  \streetaddress{15 JJ Thomson Ave.}
  \city{Cambridge}
  \country{UK}
}

\author{Dingwen Tao}
\email{ditao@iu.edu}
\affiliation{%
  \institution{Indiana University}
  \streetaddress{107 S. Indiana Ave.}
  \city{Bloomington}
  \state{Indiana}
  \country{USA}
  \postcode{47405}
  }

\author{Jiannan Tian}
\email{jti1@iu.edu}
\affiliation{%
  \institution{Indiana University}
  \streetaddress{107 S. Indiana Ave.}
  \city{Bloomington}
  \state{Indiana}
  \country{USA}
  \postcode{47405}
  }

\author{Sian Jin}
\email{sian.jin@temple.edu}
\affiliation{%
  \institution{Temple University}
  \streetaddress{1801 N. Broad St.}
  \city{Philadelphia}
  \state{Pennsylvania}
  \country{USA}
  \postcode{19122}
  }

\author{Zizhe Jian}
\email{zjian106@ucr.edu}
\affiliation{%
  \institution{University of California, Riverside}
  \streetaddress{900 University Ave.}
  \city{Riverside}
  \state{California}
  \country{USA}
  \postcode{92521}
  }

\author{Daoce Wang}
\email{daocwang@iu.edu}
\affiliation{%
  \institution{Indiana University}
  \streetaddress{107 S. Indiana Ave.}
  \city{Bloomington}
  \state{Indiana}
  \country{USA}
  \postcode{47405}
  }

\author{Md Hasanur Rahman}
\email{mdhasanur-rahman@uiowa.edu}
\affiliation{%
  \institution{University of Iowa}
  \streetaddress{201 S. Clinton St.}
  \city{Iowa City}
  \state{Iowa}
  \country{USA}
  \postcode{52246}
}

\author{Boyuan Zhang}
\email{bozhan@iu.edu}
\affiliation{%
  \institution{Indiana University}
  \streetaddress{107 S. Indiana Ave.}
  \city{Bloomington}
  \state{Indiana}
  \country{USA}
  \postcode{47405}
  }

\author{Shihui Song}
\email{shihui-song@uiowa.edu}
\affiliation{%
  \institution{University of Iowa}
  \streetaddress{201 S. Clinton St.}
  \city{Iowa City}
  \state{Iowa}
  \country{USA}
  \postcode{52246}
  }  

\author{Jon C. Calhoun}
\email{jonccal@clemson.edu}
\affiliation{%
  \institution{Clemson University}
  \streetaddress{433 Calhoun Dr}
  \city{Clemson}
  \state{South Carolina}
  \country{USA}
  \postcode{29634}
}

\author{Guanpeng Li}
\email{guanpeng-li@uiowa.edu}
\affiliation{%
  \institution{University of Iowa}
  \streetaddress{201 S. Clinton St.}
  \city{Iowa City}
  \state{Iowa}
  \country{USA}
  \postcode{52246}
}

\author{Kazutomo Yoshii}
\email{kazutomo@mcs.anl.gov}
\affiliation{%
  \institution{Argonne National Laboratory}
  \streetaddress{9700 Cass Ave.}
  \city{Lemont}
  \state{Illinois}
  \country{USA}
  \postcode{50439}
}

\author{Khalid Ayed Alharthi}
\email{kharthi@ub.edu.sa}
\affiliation{%
  \institution{University of Bisha}
  \streetaddress{Department Of Computer Science And Artificial Intelligence, College of Computing And Information Technology, University Of Bisha, Bisha 61922, P.O. Box 551, Saudi Arabia.}
  \city{Bisha}
  \country{Saudi Arabia}
}

\author{Franck Cappello}
\email{cappello@mcs.anl.gov}
\affiliation{%
  \institution{Argonne National Laboratory}
  \streetaddress{9700 Cass Ave.}
  \city{Lemont}
  \state{Illinois}
  \country{USA}
  \postcode{60439}
}

\renewcommand{\shortauthors}{Di et al.}

\begin{abstract}
Error-bounded lossy compression has been effective in significantly reducing the data storage/transfer burden while preserving the reconstructed data fidelity very well. Many error-bounded lossy compressors have been developed for a wide range of parallel and distributed use cases for years. They are designed with distinct compression models and principles, such that each of them features particular pros and cons. In this paper we provide a comprehensive survey of emerging error-bounded lossy compression techniques. The key contribution is fourfold. (1) We summarize a novel taxonomy of lossy compression into 6 classic models. (2) We provide a comprehensive survey of 10 commonly used compression components/modules. (3) We summarized pros and cons of 46 state-of-the-art lossy compressors and present how state-of-the-art compressors are designed based on different compression techniques. (4) We discuss how customized compressors are designed for specific  scientific applications and use-cases. We believe this survey is useful to multiple communities including scientific applications, high-performance computing, lossy compression, and big data.  
\end{abstract}




\maketitle

\section{Introduction}
\label{sec:introduction}

Today's scientific exploration and discovery substantially depend on large-scale scientific simulations or advanced instruments, which can easily produce vast amounts of data. Such vast volumes of data need to be transferred at different levels of devices (such as memory, network, and disk I/O) during the simulations or data acquisition for post hoc analysis. Coherent imaging methods, for example, are one of the primary drivers for the upgrades to the light sources (LCLS-II \cite{lcls-ii}, APS-U \cite{APSU}, NSLS-II, ALS-U), which 
will generate high-resolution detector images 
at a very high frequency, producing data streams of 250 GB/s \cite{cappello2019use} in some settings. 
Another typical example is 
that 16 petabytes of memory is required to store the full quantum state of a 50-qubit system \cite{fullstate_compression_quantum}. 

Lossy compression\footnote{Data compression is also known as data reduction. In this paper we may use these two terms exchangeably.} has been  effective in reducing the volumes of scientific data for different use cases. The scientific data to compress are generally referred to as the datasets used/generated by scientific applications or instruments, which are often stored in floating-point or integer values. Basically, as indicated by our previous study \cite{cappello2019use}, the common use cases that have been explored include significantly reducing storage footprint \cite{mdz} and memory footprint \cite{fullstate_compression_quantum}, avoiding recomputation cost in scientific simulations \cite{pastri}, accelerating checkpoint/restart \cite{tao2018improving}, accelerating the I/O performance \cite{8891037}, and reducing data stream intensity \cite{roibin}. More emerging use cases will be discussed in Section \ref{sec:customize}. 

Studies have showed that the reconstructed data of lossy compressors are acceptable to users for their post hoc analysis as long as the compression errors can be controlled to a certain extent. Such lossy compressors that allows to control the data distortion are often called \textit{error-bounded lossy compressors}. Since error-bounded compression can potentially reach very high compression ratios (e.g., 10--1000 \cite{Xin-bigdata18,sz3} or even higher \cite{faz, SPERR, ballester2019tthresh}), this technique is arguably a promising solution to resolve the big data issues for scientific applications. 

In this paper we present a comprehensive survey to provide a thorough understanding of the error-controlled lossy compression techniques especially for scientific datasets and how they are used in different parallel and distributed use cases. 

We identify the scientific datasets targeted here as the datasets generally represented in the form of floating-point or integer values. These datasets could be generated by scientific applications or advanced instruments (such as light-source), and they could be stored in memory for the reuse in the simulation or stored/transferred separately for posthoc analysis. Moreover, these datasets can be structured (e.g., 1D or multi-dimensional grid data) or unstructured (e.g., triangular meshes consisting of nodes, edges, and faces). Specific multi-media datasets (such as image-format or wave-format) are excluded in this survey. 

The selection of lossy compressors and research works is based on the following criteria.
\begin{itemize}
    \item These compressors are designed for scientific datasets (represented in the above-mentioned formats) generated by HPC applications or advanced X-ray facilities such as APS \cite{APSU}.
    \item These compressors provide certain error control, in the form of error bound \cite{sz16,sz17}, bit-truncation precision \cite{lindstrom2017fpzip,zfp}, or control of impact on posthoc analysis. For example, lossy compression methods such as JPEG and JPEG2000 have been used in certain applications \cite{4241350}, but they are excluded from this survey because they lack explicit mechanisms for controlling or bounding compression errors.
\end{itemize}  

The following topics are considered.

\begin{itemize}
    \item We summarize a novel taxonomy of lossy compression into 6 compression models.
    \item We provide a comprehensive survey of 10 commonly used compression components/modules (such as various predictors, bit truncation, quantization, wavelet transform, tucker decomposition, autoencoder) used in different lossy compressors.
    \item We summarize 46 state-of-the-art lossy compressors (i.e., compression pipelines) and use the representative compressors to describe how compression modules are used in the compression design. The studied compressors include not only classic general-purpose error-bounded lossy compressors (such as SZ, ZFP) but also many emerging tailored lossy compressors (such as SPERR, AESZ, FAZ, MDZ) optimized for specific use cases.
    \item We provide a comprehensive survey of many emerging parallel scientific applications and distributed use cases regarding the error-bounded lossy compression technique.
\end{itemize}   

To the best of our knowledge, this is the most comprehensive summary of the compression modules/techniques used by existing error-bounded lossy compressors (up to the year 2024), and the most comprehensive survey for the emerging state-of-the-art error-bounded lossy compressors. Moreover, we highlighted the key takeaways throughout the survey in both italic and bold font to make important insights easier to follow.

The remainder of the paper is organized as follows. We discuss related work in Section \ref{sec:related}. We propose a compression model taxonomy in Section \ref{sec:taxonomy}. In Section \ref{sec:techniques} we survey modular techniques commonly used in lossy compressors. In Section \ref{sec:generic-compressors} we discuss off-the-shelf lossy compressors for scientific datasets and how they are developed based on the aforementioned lossy compression modules/techniques. 
In Section \ref{sec:customize} we discuss a wide range of applications and the parallel and distributed use cases. We conclude the survey in Section \ref{sec:conclusion} with a discussion of future work.

\section{Related Work}
\label{sec:related}

In this section we discuss the work related to the survey of lossy compression. 

\textbf{Compression survey across domains and data types.} 
\begin{itemize}
    \item Jayasankar et al.~\cite{JAYASANKAR2021119} contributed a comprehensive survey to summarize the data compression techniques in terms of different coding schemes (such as entropy coding and dictionary coding) amd across various data types (such as text compression, image compression, audio compression, and video compression). The survey also involved various use cases, including compression for wireless sensor networks, medical imaging, database compression, HEP data compression, and wind turbine data compression.  
    \item Son et al.~\cite{son_survey} provided a survey about data compression for scientific domains that were generally produced by HPC applications. This survey involves both lossless compression techniques (such as FPC \cite{fpc}, ISOBAR \cite{isobar}, and PRIMACY \cite{primacy}) and four lossy compressors (such as ISABELA \cite{isabela} and fpzip \cite{lindstrom2017fpzip}). 
    \item There are many survey studies about image compression (to name a few \cite{icsur3,HUSSAIN201844}). These surveys focus on the compression for the image data. The major techniques include various transforms such as discrete cosine trans-
form (DCT) and discrete wavelet transforms (DWT) wavelet transform, learning-driven image compression \cite{icsur3} such as autoencoder (AE), variational autoencoder (VAE), and convolutional neural network (CNN).
\end{itemize}

The above surveys have limited information about modern lossy compression for scientific datasets. For example, they missed state-of-the-art (SOTA) error-bounded lossy compressors such as SZ \cite{sz16,sz17}, ZFP \cite{zfp}, MGARD \cite{MGARD}, SPERR \cite{SPERR}, and TTHRESH \cite{ballester2019tthresh}. Our survey provides an in-depth survey of lossy compression for scientific datasets, which features error-control or data distortion control in terms of scientific analysis. We also develop a compression model taxonomy, and present compression techniques and how lossy compressors are used in applications.

\textbf{Compression Survey for Specific Domains or Use Cases}

We also collected the survey papers about compression for specific domains, use cases, or data. 

\begin{itemize}
    \item Climate data: Mummadisetty et al.~\cite{lossless-climate} discussed the lossless compression methods used for climate datasets. This survey shows that lossless compressors can only get  compression ratios of up to 5.81 on climate data compression. Kunkel et al.~\cite{Kuhn_Kunkel_Ludwig_2016} wrote another survey about data compression for climate data, which mainly covered  lossless compression techniques and mentioned only a few  lossy compressors such as ISABELA \cite{isabela} and ZFP \cite{zfp}. It also provided a modeling for the impact of compression on performance and cost with regard to memory, I/O, and networks. 
    \item Seismic data: Hilal et al.~\cite{seismic-survey} wrote a survey about different seismic data compression methods. This survey covers multiple compression techniques, such as transformation, prediction, quantization, run length, and sampling. It can be deemed an initial attempt to provide an up-to-date overview of the research work carried out in this all-important field of seismic data processing. 
    \item Medical data: Al-Salamee et al.~\cite{medical-survey} provided a survey regarding the compression of medical image data for both lossy compression approaches (such as Fractals, wavelet, region of interest,  and non-region of interest ) and lossless approaches (such as adaptive block size, least square). Rate-distortion is considered in this survey as the main metric to investigate and evaluate the compression quality and performance. 
    \item Point cloud data: Quach et al.~\cite{quach:hal-03579360} provided a comprehensive survey about deep-learning-based point cloud compression methods. Speciﬁcally, they covered various categories of geometry and attribute compression and discussed the importance of level of detail  decomposition for compression, the limitation of separating geometry and attribute compression, and the importance of rendering in the context of compression. They also  discussed how point cloud compression relates to mesh compression and identified  their intersection.
    \item Time series data: Chiarot et al.~\cite{time-series-compression-survey} provided a survey about the principal time series compression techniques, proposing a taxonomy to classify them considering their overall approach and their characteristics. The authors also discussed the performance of the selected algorithms by comparing the experimental results that were provided in the original articles.
\end{itemize} 

\begin{wraptable}[7]{r}{.65\linewidth}
\centering
\vspace{-2mm}
\caption{Brief comparison of survey works on data compression.}
\vspace{-3mm}
\scriptsize
\begin{tabular}{|l|l|l|}
\hline
\textbf{Survey}       & \textbf{Focus}                 & \textbf{Limitation}                              \\ \hline
Jayasankar et al.     & General compression            & Lacks error-bounded compressors            \\
Son et al.            & HPC scientific data            & Lacks SOTA techniques               \\
Image Surveys         & Image compression              & Limited to image compression                \\ 
Mummadisetty et al.   & Climate data                   & Focuses on lossless methods                      \\ 
Hilal et al. & Seismic data & Limited generalizability \\
Quach et al.          & Point cloud compression        & Limited generalizability                         \\ 
\textbf{Our Survey}   & Scientific datasets            & Comprehensive, fills prior gaps                  \\ \hline
\end{tabular}
\label{tab:survey-comparison}
\vspace{0mm}
\end{wraptable}

In comparison, our survey has the following unique features: (1) a comprehensive lossy compression model taxonomy (Section \ref{sec:taxonomy}), (2) discussing modular compression techniques (Section \ref{sec:techniques}), (3) summary of 46 general-purpose error-controlled lossy compressors (Section \ref{sec:generic-compressors}), and (4) comprehensively discussing compressor customization for specific applications (Section \ref{sec:customize}). We present a comparison in Table \ref{tab:survey-comparison}.

\section{Compression Model Taxonomy}
\label{sec:taxonomy}

\begin{wrapfigure}[12]{r}{.6\linewidth}
\centering
\vspace{-2mm}
\includegraphics[scale=0.5]{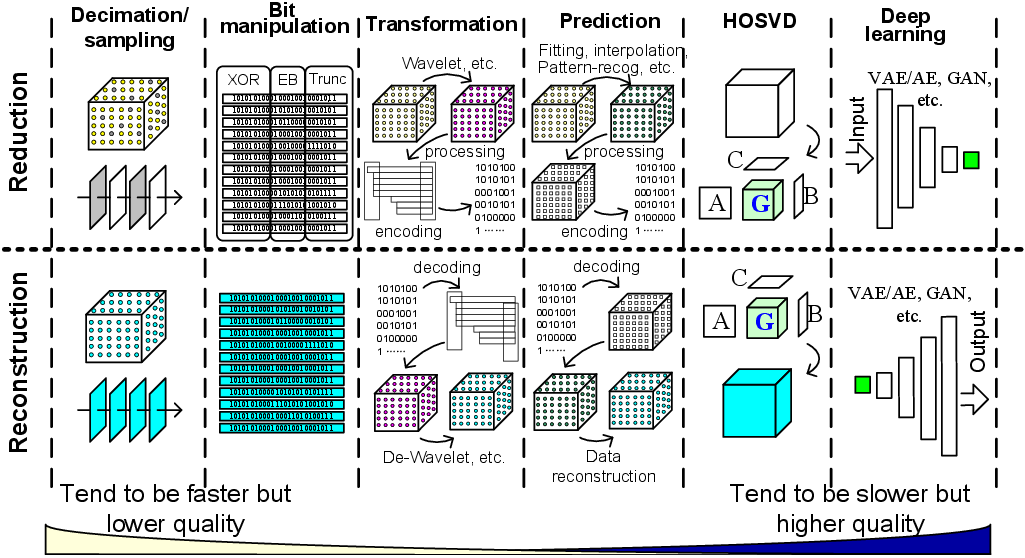}
\vspace{-8mm}
\caption{Scientific Lossy Compression Model Taxonomy}
\label{fig:comp-taxo}
\end{wrapfigure}

\begin{wrapfigure}[10]{r}{.45\linewidth}
  \centering
\vspace{-5mm}
  \raisebox{-1mm}{\includegraphics[scale=0.45]{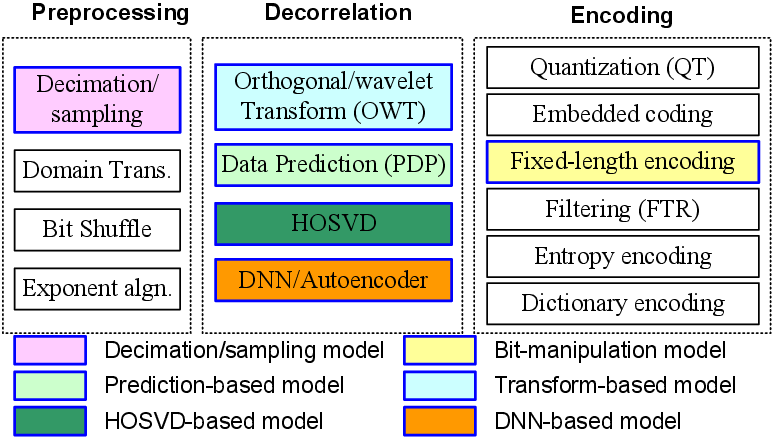}}
  \vspace{-3mm}
  \caption{Compression Pipeline with Various Models: each highlighted box represents a corresponding model. All the compression techniques shown here will be detailed in Section \ref{sec:techniques}.}
  \label{fig:taxo-steps}
  \vspace{-2mm}
\end{wrapfigure}

We developed a taxonomy derived from 46 general-purpose scientific lossy compressors (listed in Table \ref{tab:pipelines}), categorizing them into six state-of-the-art lossy compression models, as depicted in Figure \ref{fig:comp-taxo}. These six compression models are outlined using a unified compression pipeline, shown in Figure \ref{fig:taxo-steps}. The steps enclosed in blue-bordered boxes represent the most critical components of each model, which serve as the basis for naming the corresponding compression model.

\textit{\textbf{Takeaway 1}}: \textit{each model has its pros and cons with respect to the time/space complexity and reconstruction quality}. The models on the left  generally tend to have lower time complexity yet lower data reconstruction quality than do the models on the right.\footnote{Note that the performance and quality rule demonstrated in Figure \ref{fig:comp-taxo} holds in general cases. The real performance/quality also depends on the specific design and implementation.} In practice, each reduction model is a fundamental technique that can be combined with other models or techniques to generate a specific data compressor, which will be detailed in Section \ref{sec:generic-compressors}.  


In what follows, we describe the six data compression models and their pros and cons. 

\textbf{1. Decimation-/filtering-based compression}: Decimation can be split into two categories: spatial decimation and temporal decimation. The former often adopts a sampling method during the data compression and then recovers the missing data by applying an interpolation over the sampled data during the data reconstruction. The latter samples the temporal snapshots every $K$ time steps during the simulation or data acquisition and reconstructs the missing snapshots using an interpolation method. \textbf{Pros}: extremely high data compression performance. \textbf{Cons}: potentially very expensive data reconstruction since it needs to recover missing data points by numerical methods such as interpolation. Specific examples will be given in Section Section 4.7. 

\textbf{2. Bit-manipulation-based compression}: Bit manipulation is commonly used to remove the insignificant bits in the dataset, which may reduce the data size in turn. A typical example is Bit Grooming
\cite{zenderBitGroomingStatistically2016}, which analyzes the number of significant bits with respect to the user-specified number of base-2 or base-10 digits and truncates data by removing insignificant parts. Another example is SZx \cite{szx}, which pursues a very high compression speed by enforcing every step in the compression to be composed of fairly lightweight operations such as addition, subtraction, and bitwise operation. \textbf{Pros}: fairly fast data compression because of pure bitwise operations. \textbf{Cons}: relatively low data compression ratio because it does not take full advantage of the data characteristics or correlation information. 

\textbf{3. Transformation-based compression}: Data transform (such as wavelet and cosine transform) has been widely used in the data compression community \cite{zfp, vapor, vaporgithub, SPERR} because it can effectively decorrelate data by converting the original data domain to another so-called coefficient domain. The transformed domain is easier to compress as most coefficients are near zero and exhibit spatial regularity, with large values clustered at a corner of the space. For instance, the Haar wavelet in SSEM \cite{ssem} calculates deltas between adjacent data points across dimensions, often yielding many near-zero values due to the dataset's high smoothness. \textbf{Pros}:  may lead to fairly high rate distortion (i.e., high ratio with high quality); high performance due to matrix multiplication (e.g., on GPU). \textbf{Cons}: not easy to control the error bound; relatively fixed transform methods.

\textbf{4. Prediction-based compression}: A prediction-based compression model \cite{lindstrom2017fpzip,isabela,lfzip,sz16,sz17,Xin-bigdata18} generally involves four steps: pointwise data prediction, quantization, variable-length encoding, and dictionary encoding. Data prediction is the most critical step in the prediction-based compressors because higher prediction accuracy can significantly reduce the burden of the later steps. \textbf{Pros}: very high compression ratio with high quality; customizable prediction stage to fit different datasets adaptively; easy/effective control of errors. \textbf{Cons}: inferior performance (speed) because of variable-length and dictionary encoding; nontrivial to accelerate over GPUs. 

\textbf{5. HOSVD-based compression}: Higher-order singular value decomposition (HOSVD) (e.g., Tucker decomposition) can effectively decompose the data (i.e., a tensor) to a set of matrices and a small core tensor, with well-preserved L2 normal error.  By combining HOSVD and other techniques such as bit-plane, run-length, and/or arithmetic coding, the data size could be significantly reduced. \textbf{Pros}: extremely high compression ratio \cite{ballester2019tthresh} since it leverages long-range correlation in the dataset across different dimensions (such as time dimension and different fields). \textbf{Cons}: very expensive because of its intrinsic iterative steps in error control \cite{ballester2019tthresh}.

\textbf{6. Deep-learning-based compression}: Deep learning techniques have been used to improve the data compression ratio. In particular, autoencoder (AE) \cite{Goodfellow-et-al-2016} and variational autoencoder (VAE) \cite{vanillavae,dipvae,betavae,infovae,wae,swae,logcoshvae} are two classic data reconstruction techniques. An autoencoder is a kind of artificial neural network for learning efficient data codings in an unsupervised manner. 
The original aim of an autoencoder is to learn a representation (encoding) for a set of data, typically for dimensionality reduction. 
\textbf{Pros}: a fast-emerging technique with a promising opportunity to get a very high compression ratio. \textbf{Cons}: inferior data reconstruction quality;  very expensive training; relatively expensive encoding and decoding.
\section{Modular Lossy Compression Techniques}
\label{sec:techniques}

\begin{wrapfigure}[9]{r}{.2\linewidth}
  \centering
\vspace{-10mm}
  \raisebox{-1mm}{\includegraphics[scale=0.25]{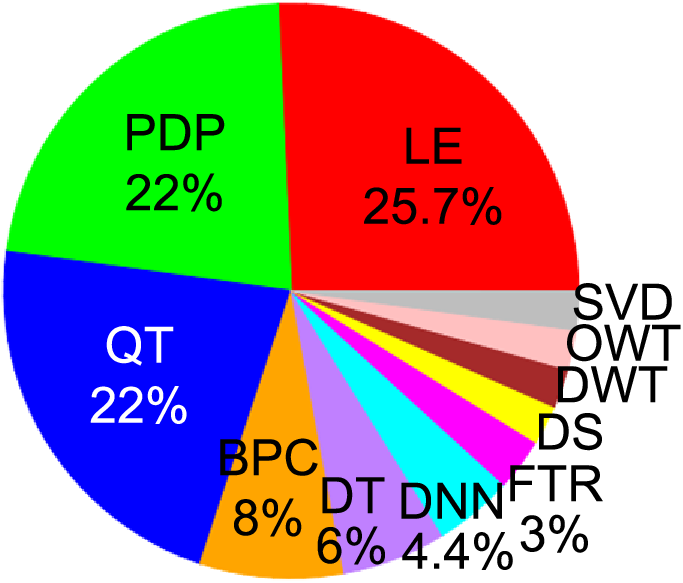}}
  \vspace{-6mm}
  \caption{Usage Distribution of Compression Techniques in 46 Compressors (e.g., 25.7\% compressors in Table \ref{tab:pipelines} used LE).}
  \label{fig:module-dist}
  \vspace{-1mm}
\end{wrapfigure}
In this section, we describe the key lossy compression modules/techniques often used in many modern state-of-the-art lossy compressors. Each technique listed here serves as just a module or step in a compression pipeline (i.e., a compressor). 

\textit{\textbf{Takeaway 2}}: \textit{each compression technique often needs to be combined with one or more other techniques to compose an error-bounded lossy compression pipeline in order to obtain a high compression ratio, forming a many-to-many relationship between compression-pipeline and compression-technique. Figure \ref{fig:module-dist} presents the usage distribution of compression techniques in the 46 lossy compressors we collected.}

\vspace{2mm}
\noindent\textbf{4.1 Pointwise Data Prediction (PDP)}

Data prediction is a critical technique in the prediction-based error-bounded compression model, such as FPZIP and the SZ-series compressors including SZ1.4, SZ2, SZ3, and QoZ, as well as many domain-specific compressors (MDZ \cite{mdz}, CliZ \cite{CliZ}, etc.). Generally in the whole compression pipeline, the data prediction step is the first or second step, followed by computing the difference between the predicted value and the original value, which would lead to a set of close-to-zero values. These close-to-zero values could be compressed more easily/effectively than the original data values. 

Two critical constraints exist in the design of the data predictor to be used in an error-bounded lossy compression model such as SZ. 

\textit{\textbf{Reconstructed-Data-Driven Policy}}. The prediction method cannot use the original raw data values directly in the course of data prediction, because the predicted values must be identical between the compression stage and decompression stage while the prediction method can see only the lossily reconstructed values during the decompression. Otherwise, compression errors cannot be bounded because of the undesired inconsistent predicted values during the compression versus decompression. 

\begin{wrapfigure}[12]{r}{.5\linewidth}
  \centering
\vspace{-5mm}
  \raisebox{-1mm}{\includegraphics[scale=0.5]{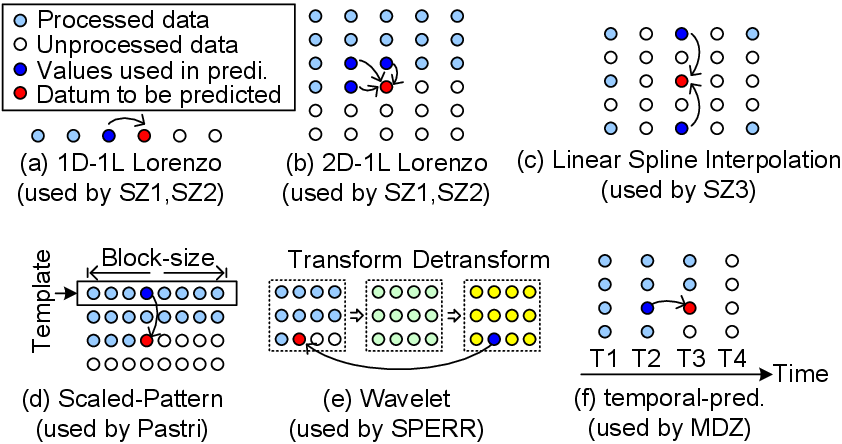}}
  \vspace{-7mm}
  \caption{RRS Policy with Six Prediction Methods. SZ1, SZ2, SZ3, SPERR are general-purpose compressors (see Table \ref{tab:pipelines}); Pastri \cite{pastri} and MDZ \cite{mdz} are customized compressors for chemistry applications (Sec. 6.1 \& 6.2). }
  \label{fig:RRS}
\end{wrapfigure}
\textit{\textbf{Recoverable Recursive-Scanning Policy (RRS policy)}}. The prediction method should be able to cover all the datapoints in terms of a specific scanning policy/order, since the data values would be reconstructed one by one in the course of decompression. Figure \ref{fig:RRS} demonstrates the RRS policy based on six eligible predictors. As shown in the figure, the scanning policy of all the prediction methods presented is executable to cover all data points throughout the whole dataset. 


Table \ref{tab:predictors} summarizes many existing predictors used in different error-bounded lossy compressors. The most popular predictors used in generic-purpose compressors include Lorenzo predictor, linear regression, spline interpolation, and wavelet transform. In general, the prediction method applied on each data point in the whole dataset leverages a certain number of neighboring or adjacent data values in spatial or temporal dimension.  


\vspace{2mm}
\noindent\textbf{4.2 Quantization (QT)}

Quantization is a popular technique widely used in today's lossy compressors. Basically, quantization means  a specific procedure/operation in which the data value range will be split into multiple consecutive intervals (i.e., quantization bins) each with a unique bin number, and each data value would be checked in which bin/interval it is located so that it can be represented by the corresponding quantization bin number. After the quantization step, each quantization bin would contain a certain number of data values, forming a histogram, which would often be encoded by a certain coding algorithm (such as Huffman encoding) to get a high compression ratio. The data to be quantized could be the original data values \cite{huang2023cuszp,zhang2023fz} or the difference of predicted value  and original data \cite{sz17,Xin-bigdata18}. We summarize various types of quantization in Table \ref{tab:quantizations}. 

\begin{wraptable}{r}{0.7\textwidth}
\vspace{-1mm}
    \caption{Predictors Used in Lossy Compressors: the last column indicates the number of neighbor data values used in that predictor. The details about the predictors can be found in the corresponding compressor papers.}  \vspace{-4mm}  
\resizebox{0.7\columnwidth}{!}{  
    \begin{tabular}{|c|c|c|c|}
    \hline
    Predictor&Compressor&Domain& \# Values Used\\
    \hline
    Lorenzo-1D-1L&SZ1-3 \cite{sz16,sz17,sz3}, FPZIP \cite{lindstrom2017fpzip}&Generic&1\\
    Lorenzo-2D-1L&SZ1-3 \cite{sz16,sz17,sz3}, FPZIP \cite{lindstrom2017fpzip}&Generic&3\\
    Lorenzo-3D-1L&SZ1-3 \cite{sz16,sz17,sz3}, FPZIP \cite{lindstrom2017fpzip}&Generic&7\\
    Mean-value&SZ2 \cite{Xin-bigdata18}&Generic&Many\\
    Linear Regression&SZ2 \cite{Xin-bigdata18}&Generic&216\\
    Linear Interpolation&SZ3 \cite{sz3}, QoZ \cite{qoz}, FAZ \cite{faz}&Generic&2\\
    Spline Interpolation&SZ3 \cite{sz3}, QoZ \cite{qoz}, FAZ \cite{faz}&Generic&4\\
    Wavelet/Orthogonal Tran.&Hybrid \cite{liang2019significantly}, SPERR \cite{SPERR}&Generic&64\\
    Scaled-Pattern&Pastri \cite{pastri} (Section 6.2)&Quantum Che.&Many\\
    Temporal Smoothness&MDZ \cite{mdz} (Section 6.1)&MD&1\\
    Multi-level&MDZ \cite{mdz} (Section 6.1)&MD&Many\\
    Mask-based&CliZ \cite{CliZ} (Section 6.4)&Climate&2 or 4\\
    \hline
    \end{tabular}}
    \label{tab:predictors}
\end{wraptable}

\textbf{Linear-scale quantization}. Linear-scale quantization is  used mainly when an error bound  needs to be respected during the compression. In this method, each quantization bin has the same length.    

\textbf{Log-scale quantization}. In log-scale quantization, the quantization bin size follows a log-scale (or exponential distribution). In general, smaller bins tend to cover denser intervals in the histogram, in order to get a balanced count distribution among all quantization bins. NUMARCK \cite{numarck} is a typical example that studied log-scale quantization. 

\textbf{Vector  quantization}. Similar to log-scale quantization, vector quantization adopts variable-length quantization bins, where the quantization bin size depends on a certain clustering (e.g., K-means) technique applied on the dataset. This can improve the data approximation accuracy when using the centroid to represent all the data values contained by the corresponding bins. Typical examples that use the vector quantization include MDZ \cite{mdz} and NUMARCK \cite{numarck}. 

\begin{wraptable}{r}{0.7\textwidth}
\vspace{-1mm} 
    \centering
    \caption{Quantizations Used in Lossy Compressors: different quantization methods feature specific approximation effect as shown in the last column.}  \vspace{-4mm}  
\resizebox{0.7\columnwidth}{!}{  
    \begin{tabular}{|c|c|c|c|}
    \hline
    Method&Compressor&Domain& Approximation Feature\\
    \hline
    Linear-scale&SZ1/2/3,etc.&Generic&Fixed-error and uneven distribution\\
    Log-scale&NUMARCK&Generic&More balanced histograms\\
    Vector-quantization&MDZ/NUMARCK&MD&Matching multilevel pattern\\
    multi-interval&Cons-SZ&Generic&Adaptation to multi-intervals\\
    \hline
    \end{tabular}}
    \label{tab:quantizations}
\end{wraptable}
\label{sec:quantization}

\textbf{Multi-interval based quantization}. In this quantization method, the quantization bins may have different lengths, depending on the user's quantity of interest on various value intervals. Thus, the multi-interval-based quantization method allows users to set different error bounds (i.e., different lengths of quantization bins) to control data distortion at different value ranges more flexibly compared with the linear-scale quantization. We refer readers to \cite{LiuHiPC21,Liu-constraints-tpds} for more details.  

\vspace{2mm}
\noindent\textbf{4.3 Orthogonal/Wavelet Transform (OWT/DWT)}

Wavelet transform, specifically the hierarchical multidimensional discrete wavelet transform, is also a useful data transform method for scientific data compression. In many cases it can effectively decorrelate and sparsify the input data to coefficients with higher compressibilities. Example wavelet transforms leveraged in existing scientific lossy compressors are the CDF9/7 \cite{CDF97} wavelet in SPERR \cite{SPERR} and Sym13 \cite{sym13} wavelet in FAZ \cite{faz}. In those compressors, the input data array is first preprocessed with wavelet transforms. Next, the transformed coefficient array is further encoded with certain encoding algorithms such as the SPECK \cite{SPECK} encoding algorithm for wavelet coefficients. The encoded bitstream usually exhibits a significantly reduced size compared with the original data.
One core limitation of wavelet transform is that, for achieving a high compression ratio, the corresponding transform often has a relatively high computational cost and therefore apparently slows  the compression process.

\vspace{2mm}
\noindent\textbf{4.4 Domain Transform (DT)}
Domain transform  here refers to a (pre)processing step that performs an operation on the dataset in order to meet a specific error bound requirement. A typical example of DT is using logarithmic domain transform to implement the pointwise relative error bound \cite{sdrb}. Specifically, Liang et al. \cite{liang2018efficient} proved that enforcing a pointwise relative error bound $e_r$ on the original data $d$ is equivalent to enforcing an absolute error bound $\log(1+e_r)$ on the logarithmic data $\log |d|$. 
Thus, compression with pointwise relative error bound can be implemented as traditional lossy compression with absolute error bound after performing a logarithmic transform on the original data. 
This is a generic approach that can be applied to any compressor with an absolute error bound. 
However, it will introduce certain overhead due to the expensive logarithmic operations during compression and exponential operations during decompression. Another good example about DT is zMesh \cite{zMesh}, which reorder the adaptive mesh refinement (AMR) dataset by an optimized space-filling curve. This can significantly improve the smoothness of the data, which thus improve the compression ratio substantially in turn.  

\vspace{2mm}
\noindent\textbf{4.5 Bit-Plane  Coding (BPC)}

Bit-plane coding  is commonly used in many lossy compressors with different compression models. BPC can be applied either in the original data domain or in the transformed coefficient domain. The fundamental idea about BPC is that the scientific data are always stored in a specific bit-plane representation (e.g., IEEE 754 floating-point or integer), such that each bit in the presentation affects the data value with different levels. 

Taking 32-bit floating-point data as an example, the alteration of leading bits (high-end) will change the data value more significantly than the alteration of ending bits (low-end). The reason is that the leading part contains the sign, exponent, and significant mantissa bits. Thus, ignoring a certain number of insignificant bit planes for a group of data values is often used in different lossy compression algorithms. In general, the loss introduced into the data by the bit truncation method is determined by the data values: the larger the data value, the larger the compression error. We explain the reason by using a floating-point value as an example. For simplicity and without loss of generality, we give the analysis based on the floating-point value in  decimal format instead of binary format actually used by the data representation on machines. For the two numbers 12.34 and 123.4, their representations are 1.234$\times$$10^1$ and 1.234$\times$$10^2$. When a digit is removed (e.g., removing 4), the errors introduced into the two numbers would be 0.04 and 0.4, respectively, which depends on the original data values.
Performing BPC after aligning the exponent of data to the same scale is a typical variation, which enforces an absolute error bound that is irrelevant to the data value.

Table \ref{tab:bit-plane} summarizes 8 error-controlled lossy compressors that involve the BPC technique. 

\begin{table}[ht]
    \centering
    \caption{Bit-Plane Coding Methods Used in Lossy Compressors: bit-plane coding is used in many compressors for various purposes, also lreadig to diverse error control modes.}  \vspace{-2mm}  
\resizebox{0.85\columnwidth}{!}{  
    \begin{tabular}{|c|c|c|}
    \hline
    Compressor&Stage \& Purpose& Error Control Mode\\
    \hline
    SZ1/2 \cite{sz16,sz17,Xin-bigdata18}&Processing Outlier/Unpredictable Data&Absolute Error Bound\\
    ZFP \cite{zfp}&Processing/Encoding Transformed Coefficients&Absolute Error Bound \& Precision Mode\\
    FPZIP \cite{lindstrom2017fpzip}&Processing Prediction-Mapped Integer Residuals& Precision Mode\\
    SZx \cite{szx}& Processing Nonconstant Blocks& Absolute Error Bound\\
    cuSZp \cite{huang2023cuszp}& Processing Nonzero Blocks after Quantization+Lorenzo&Absolute Error Bound\\
    SPERR \cite{SPERR}& Processing Wavelet-Transformed Coefficients& Absolute Error Bound\\
    DigitRounding \cite{delaunay2019evaluation}& Processing Raw Data& Absolute Error Bound\\
    BigGrooming \cite{zenderBitGroomingStatistically2016}& Processing Raw Data&Absolute Error Bound\\
    \hline
    \end{tabular}}
    \label{tab:bit-plane}
\end{table}

\vspace{2mm}
\noindent\textbf{4.6 Tucker Decomposition and HOSVD (SVD)}

Tucker decomposition, particularly  HOSVD, is a robust technique extensively utilized for data reduction in high-dimensional datasets~\cite{ballester2019tthresh, tensor-trunc-2011, tensor-trunc-2013, tensor-trunc-2015}. This method extends the matrix singular value decomposition (SVD) to higher-order tensors. HOSVD decomposes a dataset into a core tensor and a series of matrices corresponding to each dimension, effectively leveraging the spatial correlation within the dataset to capture its multidimensional structure.
As a result of the HOSVD process, the transformed core tensor becomes sparser than the original dataset, enhancing its compressibility. This characteristic enables HOSVD-based compressors to achieve significant compression ratios with minimal information loss, particularly for datasets with relatively smooth variations. The primary limitation of HOSVD lies in its computational complexity, especially for large datasets, which can make the decomposition computationally expensive and time-consuming.

\vspace{2mm}
\noindent\textbf{4.7 Decimation/Sampling (DS)}
\label{sec:decimation}

Decimation/sampling  is commonly used by scientific applications to reduce the volumes of the simulation data to be stored on parallel file systems. In general, \textit{decimation} means performing downsampling along the time dimension during the simulation: for example, saving the snapshot data to disks every $K$ time steps instead of saving all snapshots during the simulation. Many scientific simulation packages, such as Hardware/Hybrid Accelerated Cosmology Code (HACC) \cite{habib2013hacc}, EXAALT molecular dynamics simulation \cite{exaalt}, reverse time migration (RTM) \cite{rtm-analysis1,rtm-analysis2}, and Flash-X \cite{flashx}, allow users to save the snapshot data selectively over time. In comparison with decimation, the \emph{sampling} strategy generally means performing downsampling in space for each snapshot dataset, which can also significantly reduce the data volumes. Liang et al.~\cite{liangdecimation-drbsd18} studied the pros and cons of different decimation/sampling-based compression strategies in both temporal and spatial dimension. Specifically, the authors pointed out that a decimation/sampling method can have extremely high speed in the compression stage, but it may suffer from substantial decompression cost and also low reconstructed data quality compared with  traditional error-bounded lossy compressors such as SZ \cite{sz17,Xin-bigdata18}. Compressed sensing (CS) is another typical lossy compression method that leverages the sampling strategy. In general, CS is used  where the compression is required to be very fast (e.g., in  online compression) while decompression is not that important and can be performed offline. CS can be very fast  because it just needs to sample the dataset with a certain randomness. To reconstruct the data, however, CS needs to solve an underdetermined linear system, which could be very expensive. 

\vspace{2mm}
\noindent\textbf{4.8 Filtering (FTR)}

The filtering technique aims to remove insignificant values or ignore the insignificant changes of the data, which can then significantly reduce the data size. In general, the significance of the data is determined by the impact of the data being processed on the final reconstructed data quality. The filtering technique has been widely used in many existing error-bounded lossy compressors, such as (cu)SZx, cuSZp, and SPERR and the specific filtering methods often appear in different forms. In the following, we describe two forms of filtering commonly used in  lossy compressors.

\begin{itemize}
    \item \textbf{Data Folding}. Data folding aims to replace (``fold'') a set of data with one single value, provided that the variation of these data can be ignored. The error-bounded compressor SZx is a good example. SZx splits the whole dataset into many fixed-length consecutive 1D blocks. If all the data values in a block are  close to each other such that the value interval range of the block is lower than or equal to twice  the user-required error bound, then the mean of the min and max in this block can be used to replace all values in the block. Such blocks are called ``constant blocks'' in SZx. Similarly, cuSZp \cite{huang2023cuszp} also splits the whole dataset into many blocks, and each block performs uses quantization and Lorenzo prediction to decorrelate the data. Most of data values then tend to be very close to 0, and the blocks with all zeros would be just represented by a 1-byte mark. This is essentially a type of data folding.  
    \item \textbf{Data Extraction}.  Data extraction is also widely used to select the significant values or outliers from among many data points, most of which tend to be relatively small. A typical example is SZ2/SZ3, which treats the unpredictable data (data points with overlarge prediction errors compared with the predefined quantization range) as outliers and processes these outliers separately. Another example is SPERR \cite{SPERR}. SPERR adopts a SPECK algorithm, which outputs only the larger values according to a varied threshold on a set of partitioned wavelet-transformed coefficients level by level. Moreover, with the wavelet+SPECK algorithm,  some data points sill might remain whose recontructed data do not meet the user-required error bound (they are called ``outliers''). These outliers are  processed separately by SPERR, which also forms a kind of data extraction method. 
\end{itemize}

\vspace{2mm}
\noindent\textbf{4.9 Lossless Encoding (LE)}

Lossless encoding is a critical technique in error-bounded lossy compression that can help obtain a fairly high compression ratio in general because the intermediate data outputted by the previous steps tend to be very sparse. The lossless compression encoders/techniques used in the different lossy compression pipelines are summarized in Table \ref{tab:lossless} and described in detail thereafter.  

\begin{table}[ht]
    \centering
    \caption{Survey of Lossless Encoders/Compressors Used in Lossy Compression Pipelines}  \vspace{-2mm}  
\resizebox{0.8\columnwidth}{!}{  
    \begin{tabular}{|c|c|c|c|}
    \hline
    Lossless Encoder&Corresponding Lossy Compressor&Key Feature&References\\
    \hline
    Huffman Encoding (HE) & SZ1.x, SZ2.x, SZ3.x, QOZ, FAZ & Entropy Encoding & \cite{sz17,sz3,qoz,faz}\\ \hline
    Arithmetic Encoding (AE) & TTHRESH & Entropy Encoding & \cite{ballester2019tthresh}\\ \hline
    \multirow{2}*{Zlib/Zstd Encoding (ZE)} & SZ1-3, QOZ, FAZ, MGARD & \multirow{2}*{Dictionary Encoding} & \multirow{2}*{\cite{sz17,qoz,faz,digitrounding}}\\
    & Bit Grooming, Digit Rounding & & \\ \hline
    RunLength Encoding (RE) & cuSZ+, TTHRESH & Reduce Repeated Symbols& \cite{cuszplus,ballester2019tthresh}\\ \hline
    Constant-block Encoding (CE) & SZx, FZ-GPU, cuSZp & Reduce Repeated Symbols& \cite{szx,zhang2023fz,huang2023cuszp}\\ \hline
    Fixed-length Encoding (FE) & cuSZp & Fast on GPU& \cite{huang2023cuszp}\\ \hline
    Embedded Encoding (EE) & ZFP, TTHRESH & Generally Fast\&Effective& \cite{zfp,ballester2019tthresh}\\ \hline
    Predictor Encoding (PE) & SZ0.1 & Simple & \cite{sz16}\\
    \hline
    \end{tabular}}
    \label{tab:lossless}
\end{table}

\vspace{2mm}
\noindent\textbf{4.10 Deep Neural Network (DNN)}

\begin{wrapfigure}[7]{r}{.5\linewidth}
  \centering
\vspace{-5mm}
  \raisebox{-1mm}{\includegraphics[scale=0.6]{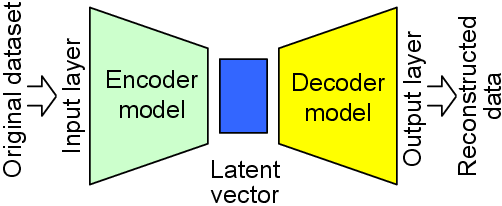}}
  \vspace{-5mm}
  \caption{Illustration of Autoencoder (AE): Encoder and decoder are two models trained with  datasets, and \textit{latent vector} represents the space in compressed format.}
  \label{fig:DNN}
\end{wrapfigure}
Since neural-network-based compression has been well developed and practicalized for natural images \cite{icsur1,icsur2,icsur3} and videos \cite{vcsur}, several initial attempts have also been made to leverage neural networks for the lossy compression of scientific data. In neural-network-based scientific lossy compressors, the neural networks can serve as both data encoders \cite{choineural,lu2021compressive,liu2021high,glaws2020deep,hayne2021using,huang2021efficient,huang2023fast,ae-sz} and data predictors \cite{huang2022compressing,han2022coordnet,KDINR,SRNN-SZ} and can also be offline-pre trained by pre-acquired datasets \cite{choineural,liu2021high,glaws2020deep,hayne2021using,huang2021efficient,huang2023fast,SRNN-SZ,ae-sz} or be online-trained by input data \cite{lu2021compressive,huang2022compressing,han2022coordnet,KDINR}. Auto-encoder (AE/VAE) can effectively reduce the dimensions or compress original data into a latent vector, as illustrated in Figure \ref{fig:DNN}. Simply using auto-encoder (AE/VAE), however, may introduce unaccepted errors \cite{liu2021high}, so many AE-based compressors leveraged AE as a predictor, followed by quantization and lossless encoding. For example, AE-SZ \cite{ae-sz} encodes the input data with a pretrained convolutional Sliced-Wasserstein Autoencoder (SWAE), and SRNN-SZ applies a pre trained Hybrid Attention Transformer (HAT) as an interpolation-like data predictor. These two compressors are typical examples with offline-trained networks. Another example (HH-NN) will be described in Section 6.4. CoordNet \cite{han2022coordnet} and KD-INR \cite{KDINR} are two compressors in which the networks are online trained by the input data before the data prediction process. Neural network-based compressors with online-trained networks can achieve much better compression ratios and/or distortions than offline-trained networks for general datsaets but suffer lower throughputs due to the requirement of training for each separate input. 




\section{General-Purpose Lossy Compressors for Scientific Data}
\label{sec:generic-compressors}

\begin{table*}
\vspace{-3mm}
    \raggedright
    \footnotesize
    \caption{Summary of General-purpose Lossy Compressors}  \vspace{-2mm}  
    \begin{adjustbox}{width=\columnwidth}
    \begin{tabular}{|c|c|c|c|c|c|c|c|c|}
    \hline
    Lossy Compressors&Year&Model&Compress. Pipeline& Device&Type&QoI&Performance & Compression Ratio\\
    \hline
    \textbf{FPZIP} \cite{lindstrom2017fpzip}& 2006 &Pred.& PDP+BPC& CPU & S &N&*** \cite{sz16} & ** \cite{Xin-bigdata18,sz16} \\ \hline
    ISABELA \cite{isabela,lakshminarasimhan2013isabela}& 2013 &Deci.& Sorting+DS+Bspline &  CPU &S&N& * \cite{di2018efficient} & * \cite{di2018efficient} \\ \hline
    \textbf{ZFP} \cite{zfp,cuZFP}& 2014&Trans.&DT+OWT+BPC& CPU/GPU&S&N& **** \cite{sz17,Xin-bigdata18} & ****\cite{sz17,Xin-bigdata18} \\ \hline    
    NUMARCK \cite{numarck}& 2014&Pred.& PDP+QT& CPU & S&N& - & ** \cite{sz16} \\ \hline
    SSEM \cite{ssem} & 2015&Trans.& DWT+QT+FTR+LE& CPU& S&N& - & ** \cite{sz16}\\ \hline
    \textbf{SZ0.1} \cite{sz16} & 2016&Pred.& PDP+LE& CPU&S&N& *** \cite{sz16} &*** \cite{sz16} \\ \hline    
    \textbf{Bitgrooming} \cite{zenderBitGroomingStatistically2016}&2016 &BitM.& BPC+LE& CPU & S&N& - & *  \cite{underwoodUnderstandingEffectsModern2022} \\ \hline
    \textbf{SZ1.4} \cite{sz17} & 2017&Pred.&  PDP+QT+LE& CPU &S&N& *** 
 \cite{sz17,Xin-bigdata18} & *** 
 \cite{sz17,Xin-bigdata18} \\ \hline
    \textbf{SZ2} \cite{Xin-bigdata18} & 2018 &Pred.&PDP+QT+LE& CPU & S&N& **** \cite{Xin-bigdata18,sz3} & **** \cite{Xin-bigdata18,sz3} \\ \hline 
    \textbf{MGARD} \cite{MGARD,ainsworth2020multilevel} & 2018 &Pred.& PDP+QT+LE & CPU &U\&S&Y& ** \cite{liang2021mgard+} & *** \cite{liang2021mgard+} \\ \hline
    \textbf{SpaioTmpDeci.} \cite{liangdecimation-drbsd18} & 2018 & Decim. & DS & CPU & S &N& **** \cite{liangdecimation-drbsd18} & * \cite{liangdecimation-drbsd18} \\ \hline
    \textbf{Digitrounding} \cite{digitrounding} & 2019 &BitM.& BPC+LE & CPU & S&N& - & * \cite{underwoodUnderstandingEffectsModern2022}\\ \hline
    DCTZ \cite{DCTZ-MSST19} & 2019 &Trans.& OWT+QT+LE & CPU & S&N& *** \cite{DCTZ-MSST19} & *** \cite{DCTZ-MSST19} \\  \hline
    GhostSZ \cite{xiong2019ghostsz} & 2019 &Pred.& PDP+QT+LE & FPGA & S&N& *** \cite{xiong2019ghostsz} & **** \cite{xiong2019ghostsz} \\ \hline 
    \textbf{TTHRESH} \cite{ballester2019tthresh} & 2019 &HOSVD& SVD+BPC+LE & CPU &S&N& * \cite{ballester2019tthresh,hpez-sigmod24,ATC23} & ***** \cite{ATC23}  \\  \hline    
    \textbf{ZFP-V} \cite{zfp-v} & 2019 &Trans.& DT+OWT+BPC+LE & FPGA & S&N& **** \cite{zfp-v} & **** \cite{zfp-v} \\ \hline    
    \textbf{SZauto} \cite{sz-auto} & 2020 &Pred.& PDP+QT+LE& CPU & S&N& *** \cite{sz-auto} & **** \cite{sz-auto}  \\ \hline    
    waveSZ \cite{wavesz}  & 2020 &Pred.& QT+PDP+LE & FPGA & S&N& *** \cite{wavesz} & **** \cite{wavesz}\\ \hline
    \textbf{TuckerMPI} \cite{tuckerMPI} & 2020 & HOSVD & SVD & CPU & S &N& ** \cite{ATC23} & ** \cite{ATC23} \\ \hline
    \textbf{zMesh}\cite{zMesh} & 2021 &Pred.& DT & CPU & U&N& *** \cite{zMesh} & **** \cite{zMesh} \\ \hline
    TEZip \cite{tezip} & 2021 & DeepL.& DNN+LE & CPU & S&N& **** \cite{tezip} & **** \cite{tezip}\\ \hline
    cuSZ \cite{cusz,cuszplus} & 2021 &Pred.& QT+PDP+LE& GPU & S&N& ** \cite{huang2023cuszp}& *** \cite{huang2023cuszp,huang2024cuszp2,cuszi} \\ \hline
    \textbf{MGARD+} \cite{liang2021mgard+} & 2021 & Pred. & PDP+QT+LE & CPU & S &N& *** \cite{liang2021mgard+} & **** \cite{liang2021mgard+} \\ \hline
    \textbf{SZ3} \cite{sz3,sz3modular} & 2021 &Pred.& PDP+QT+LE & CPU & S&N& *** \cite{sz3} & **** \cite{sz3,hpez-sigmod24,faz}\\ \hline
    \textbf{AESZ} \cite{ae-sz,liu2021high} & 2021 &DeepL.& DNN+PDP+QT+LE & CPU & S&N& ** \cite{ae-sz} & **** \cite{ae-sz} \\ \hline
    CAE \cite{liu2021high} & 2021 &DeepL.& DNN+PDP+QT+LE& CPU & S&N& ** \cite{liu2021high} & **** \cite{liu2021high} \\ \hline
    \textbf{MGARD-GPU}\cite{chen2021accelerating, gong2023mgard}& 2021 &Pred.& PDP+QT+LE & GPU & S&N& *** \cite{gong2023mgard} & *** \cite{gong2023mgard} \\ \hline
    \textbf{DE-ZFP}\cite{de-zfp} & 2022 &Trans.& DT+OWT+BPC+LE& FPGA &S&N& **** \cite{de-zfp} & **** \cite{de-zfp} \\ \hline 
    \textbf{QoZ} \cite{qoz} & 2022 &Pred.& PDP+QT+LE& CPU & S&N& *** \cite{qoz,hpez-sigmod24} & **** \cite{qoz,hpez-sigmod24}\\ \hline      
    \textbf{SZx} \cite{szx} & 2022 &BitM.& FTR+BPC& CPU & S&N& ***** \cite{szx} & ** \cite{szx} \\ \hline
    cuSZx \cite{szx} & 2022 & BitM. & FTR+BPC & GPU & S&N& *** \cite{huang2023cuszp} & ** \cite{huang2023cuszp} \\ \hline
    Jiao et al. \cite{JiaoD00TT0C22} & 2022 & Pred. & DT+PDP+QT+LE & CPU & S&Y& ** \cite{JiaoD00TT0C22} & *** \cite{JiaoD00TT0C22} \\ \hline
    \textbf{MGARD-Lambda} \cite{mgard-lambda} & 2022 &  Pred. & PDP+QT+LE  & CPU & S & Y& - & - \\ \hline
    ATC \cite{ATC23} & 2023 & HOSVD & SVD+QT+BPC+LE& CPU &  S&N& ** \cite{ATC23} & ***** \cite{ATC23} \\ \hline
    \textbf{ZFP-X} \cite{zfp-x} & 2023 &Trans.& DT+OWT+BPC+LE & CPU & S&N& **** \cite{zfp-x} & **** \cite{zfp-x}\\ \hline
    \textbf{SPERR} \cite{SPERR} & 2023 &Trans.& DWT+QT+LE& CPU & S&N& ** \cite{SPERR,faz} & ***** \cite{SPERR,faz}\\ \hline
    \textbf{HH-NN} \cite{hh-nn} & 2023 & DeepL.& DS+DNN+QT & GPU & S&N& * \cite{hh-nn} & ***** \cite{hh-nn}\\ \hline
    \textbf{FAZ} \cite{faz} & 2023 &Pred.& DWT+PDP+QT+LE& CPU & S&N& ** \cite{faz} & ***** \cite{faz}\\ \hline
    FZ-GPU \cite{zhang2023fz} & 2023 &Pred.& QT+PDP+LE& GPU & S&N& **** \cite{zhang2023fz} & ** \cite{zhang2023fz} \\ \hline
    \textbf{cuSZp} \cite{huang2023cuszp,huang2024cuszp2} & 2023 &Pred.& QT+PDP+LE& GPU & S&N& ***** \cite{huang2023cuszp,huang2024cuszp2} & ** \cite{huang2023cuszp,huang2024cuszp2} \\ \hline     
   \textbf{ AMR-Comp} \cite{Wang_hpdc2022,Wang_tpds2024,amric,Wang_drbsd,wang2024high}& 2023 &Pred.& PDP+QT+LE& CPU & U&N& *** \cite{Wang_tpds2024} & **** \cite{Wang_tpds2024} \\ \hline
   \textbf{ SRNN-SZ} \cite{SRNN-SZ}& 2023 &Pred.& DNN+PDP+QT+LE & CPU & S&N& * \cite{SRNN-SZ} & ***** \cite{SRNN-SZ}\\ \hline
    SZ\_ADT \cite{sz-adt} & 2023 &Pred.& PDP+QT+LE & CPU & S&N& *** \cite{sz-adt} & **** \cite{sz-adt}\\ \hline
    \textbf{HPEZ(QOZ2)} \cite{hpez-sigmod24} & 2024 & Pred. & PDP+QT+LE & CPU & S &N& *** \cite{hpez-sigmod24} & **** \cite{hpez-sigmod24}\\ \hline
    \textbf{HAE} \cite{hae} & 2024 & DeepL. & DT+DNN+QT+FTR & GPU & S &N& - & **** \cite{hae} \\ \hline
    SZp \cite{szp,hzccl} & 2024 & Pred. & QT+PDP+LE & GPU & S &N& ***** \cite{hzccl} & ** \cite{hzccl} \\ \hline
    \end{tabular}
    \end{adjustbox}
    \footnotesize \textbf{Footnote}: For the type column: S means Structured, U means Unstructured. Perf\&Ratio column gives a rough ranking based on literature: more `*' means higher throughput or ratio). The QoI column indicates whether the compressor supports preservation of Quantity of Interest (QoI). `-' means no published sources to indicate its performance or compression ratio. 31 compressors, highlighted in bold font, are described more or less in the following text.
    \label{tab:pipelines}
    \vspace{-1mm}
\end{table*}

In this section, we describe general-purpose lossy compressors, each of which projects a specific compression pipeline composed of a series of steps or techniques/modules. 

Table \ref{tab:pipelines} summarizes 46 lossy compressors, each with a distinct design or compression pipeline. In the last two columns of Table \ref{tab:pipelines}, we present an examination of the performance (compression/decompression speed) and efficiency (compression ratio) for all the listed compressors, based on published results. Performance and compression ratio are categorized into five levels, with more stars indicating higher levels. The levels are also determined considering the device and the comparison among other works. For example, SPERR, HH-NN and FAZ have the top level of compression ratio (*****) because their compression ratios can be 1-2 orders of magnitude higher than that of other state-of-the-art compressors like SZ and ZFP. CuSZp is ranked as the top level of the performance because it features a very high end-to-end compression/decompression throughput on GPU (200-400 GB/s on a single A100 GPU), which is significantly higher than other GPU compressors do \cite{huang2024cuszp2}. Note that this rating is a rough estimation based on the literature: e.g., the compression ratio level * means 1$\sim$5 while the level ***** means one thousand or even higher at the similar level of reconstructed data quality (in terms of PSNR or visualization). 

\textbf{\textit{Takeaway 3}}: \textit{We summarize the key insights based on Table \ref{tab:pipelines} as follows. (1) Development of new lossy compressors evolves very quickly especially after 2021. (2) A large majority of lossy compressors (especially for the modern ones) are composed of multiple techniques. (3) CPUs are more commonly used for compressor development than GPUs and FPGAs, likely due to the stricter design requirements and more complex programming models associated with GPU and FPGA development. (4) There are far fewer compressors for unstructured data over structured data, due to the more complex processing and decorrelation required for unstructured data. (5) Compression ratio and performance form a non-trivial tradeoff. If users want to choose a compressor that meets their performance or compression ratio requirements, we highly recommend testing those rated with at least three stars in the list. }

\begin{wraptable}[11]{r}{.45\linewidth}
\centering
\caption{Exemplifying multiple applications are tested by general-purpose compressors based on \cite{hpez-sigmod24,SPERR,zfp,ballester2019tthresh} (MD: Molecular Dynamics,  COM: Combustion, COS: Cosmology, CL: Climate, CH: Chesmitry, TU: Turbulence, GE: Geology, FU: Fusion.)}
\vspace{-3mm}
\scriptsize
\begin{tabular}{|l|l|}
\hline
\textbf{Compressor} & \textbf{Application Datasets Tested}                              \\ \hline
SZ1/2/3  & MD, COM, COS, CL, CH, TU, GE, FU, etc.   \\
ZFP & COM, COS, CL, CH, GE, TU, etc.\\
MGARD & COM, COS, CL, CH, GE, TU, etc.\\
TTHRESH & COM, CL, GE, TU, etc.\\
SPERR & COM, COS, CL, CH, GE, TU, etc. \\ \hline
\end{tabular}
\label{tab:apps}
\vspace{-1mm}
\end{wraptable}
In the following text, we use some representative state-of-the-art compressors to describe the compression design/pipelines and unique features. The lossy compressors selected here for further description  represent typical models/families in lossy compression and they are featured in the ``Compression for Scientific Data'' tutorial at SC24 \cite{sc24-tutorial}. Many of them have been widely used in the community and frequently evaluated in lossy compression and compressibility analysis studies, including \cite{sdrb,lu2018understanding,Xin-bigdata18,underwood2023black,hasan-msst24,fxrz,CAROL,cuszi}. We exemplify some example scientific applications used to evaluate different compressors (identified by existing publications \cite{hpez-sigmod24,SPERR,zfp,ballester2019tthresh}), in Table \ref{tab:apps}.

\vspace{2mm}
\noindent\textbf{SZ-family compressors.} SZ \cite{sz16,sz17,sz3,sz3modular,sz-hardtocompress-tpds} is a prediction-based error-bounded lossy compression framework. SZ's compression pipeline is composed of four stages: pointwise data prediction, quantization, variable-length encoding and lossless encoding. For different domain datasets and use cases, many predictors such as Lorenzo, linear regression, dynamic spline interpolation \cite{sz3}, have been developed. In addition to the data prediction stage, SZ developers explored improving compression capability by other stages (e.g., different quantization methods  \cite{LiuHiPC21,Liu-DRBSD21,Liu-constraints-tpds}). SZ adopts Huffman+Zstd \cite{sz16,sz17,Xin-bigdata18} to compress the quantization bins, which can significantly reduce the data size \cite{Liu-study-lossless-bigdata18}. SZ also supports pointwise relative error bound compression  \cite{liang2018efficient}, by leveraging a preprocessing step to transform the original data domain to the logarithm domain. To accelerate the throughput, an efficient fusion of logarithm was developed in \cite{Zou-MSST19,Zou-TPDS20}. 

Based on SZ3 framework, SZ team developed many variants to adapt to diverse user-demands. For example, QoZ (quality-oriented compressor) \cite{qoz} supports compression autotuning according to user-specified quality metric targets. QoZ2.0/HPDZ \cite{hpez-sigmod24} brings several major updates to its data prediction design, including multidimensional interpolation, interpolation re-ordering, dimension autofolding, and blockwise interpolation tuning, which can achieve 50\% to 300\% compression ratio improvement. FAZ \cite{faz} combines SZ3 model with wavelet transform, also adaptively leveraging the best-fit compression techniques for each separate input and autodetermine their corresponding parameters. FAZ features extremely high compression ratios over SZ3 ( at a cost of compression throughput): it can get 3-4$\times$ compression ratio as high as that of QoZ on the SEGSalt dataset as shown in \cite{faz}. In addition, the SZ team also explored leveraging DNN to improve compression ratios under the SZ3 framework: e.g., AE-SZ \cite{ae-sz} leverages AutoEncoder and SRNN-SZ \cite{SRNN-SZ} adopts super-resolution technique. In addition, the SZ team also developed several ultra-fast versions for both CPU and GPU devices. SZx \cite{szx}, for example, composes only lightweight operations, such as bitwise operations, additions, and subtractions, so that it can get a very high compression/decompression speed (about 4$\times$ as fast as SZ2 and ZFP in general). CuSZp \cite{huang2023cuszp, huang2024cuszp2} can get a very high end-to-end performance on GPU for both compression and decompression (200-400GB/s on A100 in general), still maintaining a relatively high compression ratio.

\vspace{2mm}
\noindent\textbf{ZFP-family compressors.}
ZFP \cite{zfp} is a transform-based error-bounded lossy compressor, which supports two error control methods: fixed-accuracy (i.e., absolute error bound) and fixed precision. ZFP splits the whole dataset into many fixed-size blocks (e.g., 4$\times$4$\times$4 for a 3D dataset) and then executes three steps in each block: (1) preprocessing (DT): align the values in
a block to a common exponent and convert the floating-point values to a fixed-point representation; (2) (near)orthogonal block transform (OWT): use orthogonal transform to decorrelate data; and (3) embedded coding (BPC): order and encode the transform coefficients by the embedded coding. 
To achieve the best trade-off between decorrelation efficiency and speed, developers of ZFP explored multiple transforms using a parametric description and identified a near-orthogonal one to use in practice. 
Their embedded encoding is a variation of BPC, where the coefficients are divided into separate groups based on their locations and then encoded in the granularity of a group. 
In general, ZFP features high throughput both both compression and decompression because of the performance optimization strategies in its implementation, such as lifted transform. ZFP also features multiple variants to suit different devices or use-cases: e.g., ZFP-V and DE-ZFP are two hardware-optimized lossy compressor developed based on ZFP design principle; ZFP-X is an accelerated version with remodeled embedded coding. 

\vspace{2mm}
\noindent\textbf{MGARD-family compressors.}
MGARD \cite{MGARD, ainsworth2019multilevel, ainsworth2019multilevel-qoi, ainsworth2020multilevel, gong2023mgard} is a multilevel data compressor based on finite element analysis and wavelet theories. 
It treats the data as a piecewise multilinear function defined on the input data grid and iteratively decomposes the data into coarse representations in a set of hierarchical grids. 
The decomposition procedure is  as follows. 
Starting with the original data and input grid, MGARD will compute the piecewise linear interpolation using data from the lower-level grid and then subtract the interpolation values from current data to obtain multilevel coefficients. 
These coefficients are then projected to the lower-level grid to compute correction, which roughly approximates the loss of missing nodes using the lower-level grid. 
The correction then is added to data in the lower-level grid to form the lower-level representation. 
This process is repeated until the lowest level is reached. All the multilevel coefficients are then fed to a Huffman encoder and a lossless encoder for size reduction. MGARD can be applied to uniform/nonuniform structured and unstructured grids~\cite{ainsworth2020multilevel} because of the general data decomposition theory. 
In addition to providing general error controls (such as absolute error and $L^2$ error) on raw data, MGARD features error controls on derived quantities such as bounded-linear analysis~\cite{ainsworth2019multilevel-qoi}, and also non-linear functions in MGARD-Lambda version \cite{mgard-lambda}. It also provides error control for more complex derived quantities using a postprocessing method~\cite{lee2022error, banerjee2022algorithmic}. 
The performance of MGARD is slightly slower than that of SZ and ZFP due to the higher computational complexity, but it provides an accelerated version for structured dataset (MGARD+ version \cite{liang2021mgard+} and a portable GPU implementations across different vendors with high performance.  

\vspace{2mm}
\noindent\textbf{SPERR.}
SPERR \cite{SPERR} is a transform-based lossy compressor based on the CDF9/7 discrete wavelet transform \cite{CDF97} and SPECK encoding algorithm \cite{SPECK}, and it has both a pointwise error-bounding mode and a global quality thresholding mode. 
The compression pipeline of SPERR includes four stages: (1)) CDF9/7 wavelet transform; (2)) SPECK lossy encoding of wavelet coefficients; (3)) outlier encoding (only in error-bounding mode); and (4) zstd 
postprocessing of compressed data (optional). The decompression pipeline is an inverse of the compression pipeline with the decoding, detransform, and so on. 
The advantage of SPERR is that the hierarchical multidimension DWT in SPERR can effectively capture the relevance between data points, and it can often decorrelate the transformed coefficients to a great extent, which also brings a high compression ratio after the SPECK encoding. One limitation of SPERR is that the wavelet transform and the SPECK encoding processes have high computational costs,and hence its (sequential) execution speed is relatively low, typically around 30\% of SZ3 \cite{sz3}.

\vspace{2mm}
\noindent\textbf{TTHRESH.}
TTHRESH \cite{ballester2019tthresh} is a lossy compressor that utilizes the Tucker decomposition, specifically higher-order singular value decomposition. Unlike other lossy HOSVD-based compressors \cite{tensor-trunc-2011, tensor-trunc-2013, tensor-trunc-2015}, which implement coarse-granularity slicewise truncation on the tensor core and factor matrices post-HOSVD, TTHRESH employs bit-plane coding across the entire set of HOSVD transform coefficients. This approach is complemented by run-length encoding (RLE) and arithmetic coding (AC). Notably, TTHRESH is capable of achieving significantly higher compression ratios compared with other compressors, especially for larger error bounds (i.e., higher compression ratios). This superior performance is largely attributable to HOSVD's efficiency in capturing the global correlations within the dataset. However, TTHRESH exhibits much lower speed compared with other compressors due to the high computational complexity of HOSVD; for instance, it is $O(n^4)$ for a 3D dataset with dimensions of $n^3$. Additionally, we note that TTHRESH does not offer pointwise error control; instead, it can control only the $l^2$ error (the sum of squared errors) because of the nature of HOSVD.

\vspace{2mm}
\noindent\textbf{FPZIP.}
FPZIP \cite{lindstrom2017fpzip} is an error-controlled lossy compressor developed based on the prediction-based compression model. It involves four steps: (1) It uses a Lorenzo predictor to predict the data value for each data point. (2) It computes the prediction residuals and maps these to integers. (3) After the mapping, a two-level compression scheme is applied on the residual integers. (4) It then applies a fast entropy coding (arithmetic coding) to improve the compression ratio. FPZIP does not support absolute error bound or decimal digit control, but it allows users to specify the number of bit planes (i.e., precision) to ignore, based on which the users can control the data distortion on demand. Specifically, when the precision is set to 32 for the single-precision floating-point dataset,  all  32 bit-planes will be preserved, projecting essentially a lossless compression. The lower the precision value, the higher the data distortion and also the higher the compression ratio.

\vspace{2mm}
\noindent\textbf{HAE.} The Hierarchical AE-based Lossy Compressor (HAE) \cite{hae} is a representative compressor designed using an Auto-Encoder (AE) to handle scientific datasets. As noted earlier, modern lossy compressors often integrate multiple compression techniques to achieve high compression ratios and quality, and HAE is no exception. It consists of four main steps: (1) data standardization, (2) encoding and decoding using a hierarchical Auto-Encoder, (3) quantization of the latent vector, and (4) error-thresholding of the reconstructed data.
The data standardization step serves as a form of ``domain transformation," converting each original data point into a standardized value as a preprocessing step for compression. The hierarchical Auto-Encoder reduces the data dimensionality iteratively, followed by quantization. Finally, the error-thresholding step identifies ``outlier" data points that exceed user's error bound and preserves their original values to maintain accuracy. 

\vspace{2mm}
\noindent\textbf{Digit Rounding and Bit Grooming.} 
Digit Rounding \cite{digitrounding} and Bit Grooming \cite{zenderBitGroomingStatistically2016} are two error-bounded lossy compressors, which both mainly adopt the bit-plane  coding  method. We describe these two compressors in detail as follows.

Digit Rounding  allows users to specify a decimal digit (denoted as \textit{nsd}) to preserve for the compression. For example, if a user sets the nsd to be 4 to compress the number 3.14159265, then four significant digits will be preserved: the lossily reconstructed number would be 3.14111328. Digit Rounding includes three key steps: 
\begin{itemize}
    \item Bit truncation: computing the required number of bits to preserve in the IEEE-754 floating-point representation according to the number of significant decimal digits specified by the user (i.e., nsd).
    \item Shuffle: applying a byte shuffle function on the bit-truncated dataset. 
    \item Lossless compression: compressing the shuffled bytes with a lossless compressor such as Deflate (Gzip \cite{gzip}) or Zstd \cite{zstd}). 
\end{itemize}
The official release of Digit Rounding \cite{digitrounding} has a dependency on HDF5 because it uses the deflate function offered by HDF5. 

Bit Grooming  is developed mainly based on the bit plane encoding. Similar to Digit Rounding, Bit Grooming also truncates the bit planes for the floating-point datasets by removing the insignificant digits, followed by a deflate lossless encoder such as zlib \cite{zlib}.
Bit Grooming was released together with NetCDF operators (NCOs) \cite{ncos}, so its installation depends on the NCOs package.

\section{Customized Compressors for Specific Applications or Use Cases}
\label{sec:customize}

In this section, we comprehensively survey error-bounded lossy compressors specifically tailored to particular applications and use cases. We include these applications in our survey primarily because the associated compression methods explicitly support error bounding or preserve quantities of interest (QoIs), thus ensuring fidelity relevant to their respective domains.

\textbf{\textit{Takeaway 4}}: \textit{Each use-case or application projects non-trivial requirements on posthoc analysis, compression ratio or performance, which often needs a codesign with corresponding domain experts. }

\vspace{2mm}
\noindent\textbf{6.1 Compression for Molecular Dynamics Simulations}

Molecular dynamics (MD) simulations have become one of the most important research methods in many science domains, including physics, biology, and materials science. 

The volume of data generated by MD simulations is growing exponentially, and it becomes a critical challenge for researchers to keep all of the data in their storage facilities. For instance, running MD simulations to model the SGLT membrane protein may take $2.4 \times 10^8$ steps (480 ns), resulting in approximately 260 TB of raw trajectory data with only 90,000 particles \cite{hrtc}. On the other hand, a 20-trillion particle simulation \cite{MDSim-20trillion} may produce petabytes  of data with just 10 steps. 

Lossy compression has been widely considered a promising solution to reduce the data volume of MD simulations. 
There is ongoing research into lossy compression methods tailored for MD simulations. The HRTC method \cite{hrtc} employs a strategy that represents trajectories as piecewise linear segments, coupled with quantization that is controlled for errors and a representation using variable-length integers. The PMC approach \cite{pmc} leverages the information about atomic bonds within molecules to forecast the positions of atoms in each frame; however,  this technique does not apply to simulations involving nonbonded interactions.
Omeltchenko et al.~\cite{cpc2000} proposed a spatial compressor for MD datasets that includes three steps: (1) converting all floating-point values (both position and velocity) to integer numbers, (2) building a uniform oct-tree index according to the space-filling curve of the position fields, and (3) sorting the particles based on R-indices using a radix-similar sorting method in each block and encoding the difference in adjacent indices by variable-length encoding. Tao et al.~\cite{tao2017depth} improved Omeltchenko et al.'s method by sorting the particles based on a partial-radix sorting algorithm while preserving the same compression ratios by using SZ to compress the reordered coordinates instead of directly using the R-index. 
Essential dynamics \cite{ED} is a powerful analysis tool for identifying the nature and relative importance of the essential deformation modes of a macromolecule from MD samplings. ED offers lossy compression by adopting principal component analysis (PCA) over the full trajectories of all particles. By comparison, Kumar et al.~\cite{pcadct} combined PCA and discrete cosine transform (DCT) to compress the full trajectories of the total MD dataset. Such full-trajectory-based compression methods, however, may be impractical for many of today's large-scale MD simulations. Zhao et al.~\cite{mdz} proposed an SZ2-based compressor, MDZ, which is equipped with spatial-clustering-based prediction and two-level temporal prediction. MDZ achieves high compression ratios particularly on MD simulations focusing on crystalline materials or featuring a continuous temporal domain. 

\vspace{2mm}
\noindent\textbf{6.2 Compression for Quantum Chemistry Simulations}


Quantum chemistry applications may produce extremely large amounts of data (such as petabytes of data \cite{pastri}) during  execution on parallel systems. General Atomic and Molecular Electronic Structure System (GAMESS) \cite{gamess} is a typical example. In GAMESS, the Schr\"odinger differential equation needs to be solved to obtain the wavefunction that contains all the information about a chemical system. The most expensive step in this procedure involves computation of two-electron repulsion integrals (ERIs), which takes  about 87\% time of Hartree--Fock computation time in GAMESS \cite{pastri}. This step also projects a high storage requirement because it scales as $O$($N^4$) with the size of the chemical system. ERIs are required by each time step during the simulation, but they cannot always be kept in memory because of limited memory capacity, so they need to be recomputed from scratch at every iteration. 

An error-bounded lossy compressor called Pattern Scaling for Two-electron Repulsion Integrals (PaSTRi) was developed for GAMESS, to avoid such an expensive ERI recomputation cost. Specifically, PaSTRi was developed based on the prediction-based compression model (similar to SZ). The key advantage of PaSTRi is that it leverages the inherent scaled repeated pattern features in the ERI datasets to significantly improve the prediction accuracy, which thus can considerably improve the compression ratio in turn. According to \cite{pastri}, PaSTRI exhibits much higher compression ratios than the general-purpose compressors  SZ and ZFP with different error bound settings. For example, SZ and ZFP can get  compression ratios of 7.24$\times$ and 5.92$\times$, respectively, on the compression of double-precision floating-point ERIs data, respectively, when the error bound is set to $10^{-10}$. In comparison, PaSTRi can get the compression ratio up to 16.8$\times$. Experiments also show that the performance of retrieving ERIs can be improved 200--300\% with PaSTRi over the traditional ERIs recomputation method, when the same integral data needs to be used for a total of 20 iterations during the simulation.  





\vspace{2mm}
\noindent\textbf{6.3 Compression for Quantum Circuit Simulations}

Quantum circuit simulation is employed for a variety of quantum computing research tasks, including the development of new quantum algorithms, co-design of quantum computers, and verification of quantum supremacy claims \citep{preskill2012quantum}. With limited access to today’s noisy intermediate-scale quantum \citep{preskill2012quantum} devices as well as limited performance of these devices, quantum circuit simulation on classical computers can serve as a pragmatic tool for researchers exploring these tasks. Two important types of quantum circuit simulation are Schr\"{o}dinger algorithm full state vector simulations \citep{Raedt2006MassivelyPQ}\citep{smelyanskiy2016qhipster} and tensor network contractions \citep{markov_tensor_network}. 

Full state vector simulations involve storing a quantum state vector in memory and evolving the state vector with gates over each time step. For these simulations, the space complexity scales exponentially with the number of qubits and polynomially with the circuit depth.  As circuits for simulation grow in complexity,  in terms of both number of qubits and depth of circuit, serious computational and memory limitations emerge. In order to precisely simulate the evolution of a complete $n$-qubit state vector, $2^{n}$ state vector amplitudes must be stored. Assuming complex, single-precision floating-point values are stored for each amplitude, the Frontier supercomputer, with ~4.8 PB of memory \citep{frontier_user_guide}, would be capped at a 49-qubit simulation. Today’s quantum devices are already exceeding this number of qubits, such as IBM’s Osprey with 433 qubits \citep{IBM_newquantum}.

Tensor network contraction simulators represent a quantum circuit as a tensor network, where quantum gates or states are represented as a tensor \citep{markov_tensor_network}. Indices represent the index of a bitstring that a gate operates on. Contracting tensors requires multiplication of tensors and a summation. Tensor networks can require up to the same level of memory as full state vector simulations, depending on the circuit. Even with lower-memory footprint circuits, tensor networks can have tensors grow larger and larger as the contraction sequence advances, straining the memory resources.

Compression is an attractive solution to this memory footprint problem. With sufficiently high throughput compression, large state vectors can be stored in memory and processed in chunks, eliminating the need to read and write from storage. 

Wu et al.~\citep{fullstate_compression_quantum} designed a Schr\"{o}dinger algorithm-based full state vector simulation pipeline that integrates compression. 
MPI is used to parallelize the matrix multiplication required to apply a gate to a state vector. Each rank stores a set of compressed blocks that together compose a component of the overall state vector. 
Wu et al. explored multiple compressors to compress the state vector data, including SZ2.1 (A), SZ2.1 with complex type support (B), XOR leading-zero reduction coupled with bit-plane truncation and zstd (C), and reshuffling of real and imaginary parts together before performing C (D). The results indicate that solutions C and D achieve the highest compression ratios (30--90 for error bounds in the range [1E-5,1E-1]).
When running a 61-qubit Grover’s search algorithm with this compressor, the memory requirement drops from 32 exabytes to 768 terabytes using 4,096 nodes. In all, their method can raise the number of qubits for a simulation by 2 to 16.

Shah et al.~\citep{shah_gpu_qc_compression} targeted tensor network-based quantum circuit simulation and proposed a GPU-based compression framework for these types of simulators. Since the target is tensors that exhibit spiky behavior, the authors applied preprocessing and postprocessing steps to cuSZ and cuSZx, the GPU implementations of SZ and SZx. Additionally, the cuSZx kernel was modified to integrate the pre- and postprocessing such that the impact on throughput was limited. At a high level, the pre- and postprocessing sparsify the tensor, leveraging the fact that many tensor values are close to zero and have little impact on the contraction result. 
Their designs can yield up to 10 times greater compression ratio compared with cuSZ alone. When prioritizing throughput, the modified cuSZx kernel compressor can achieve 3 to 4 times improvement in compression ratio with limited impact on throughput.

\vspace{2mm}
\noindent\textbf{6.4 Compression for Climate Research}

The simulations used by climate scientists produce enormous volumes of data.
For example, the Coupled Model Intercomparison Project alone produced nearly 2.5 PB of data \cite{cinquiniEarthSystemGrid2014}, and future studies will produce more data as  the resolution of the modeling is increased.
The data from these studies are often  extensive and used as the baseline for studies of different aspects of climate science.
Thus, both the quality and size of these datasets are of utmost importance.

Climate researchers have developed some of the most extensive work \cite{bakerEvaluatingLossyData2016,Baker-Climate17,bakerDSSIMStructuralSimilarity2022,gmd-17-8909-2024} to quantify the impacts of lossy compression on their specific domain quantities of interest.
 In a series of papers researchers proposed four critical assessments: the SSIM of the visualization (and later the data with dSSIM \cite{dssim}), the p-value of the KS test, the Pearson correlation coefficient of determination, continuous ranked probability score (CRPS) \cite{gmd-17-8909-2024}, and the spatial relative error with corresponding thresholds to be established by asking a panel of domain experts if the data were distinguishable from the datasets \cite{bakerEvaluatingImageQuality2019}.
These thresholds were later refined by the community.
Work by Underwood and Bessac \cite{underwoodUnderstandingEffectsModern2022} identified several weaknesses in the p-value of the KS test when used in this way, making the test too conservative in some cases and too liberal in others. 
Therefore, they proposed some alternative distance measures for the climate community to consider.
Additionally, some papers have proposed less aggressive limits for the SSIM metric in particular \cite{klowerCompressingAtmosphericData2021}.


As for the compressors customized for climate datasets, two typical examples include CliZ \cite{CliZ} and Huang et al. \cite{hh-nn}. 
CliZ \cite{CliZ} is a kind of prediction-based compressor, which adopts multiple optimization strategies by leveraging climate data properties such as mask-map, periodicity and geographic consistency. It also features some advanced techniques such as dimension permutation and fusion, which can further help increase prediction accuracy. Some climate datasets, like those used in Land or ICE, 
have many small fields and would benefit from common dictionary encoding optimizations. Experiments show that CliZ outperforms SZ3, SPERR, or QoZ1.1 on climate datasets by 20\%-200\% in compression ratio with comparable time cost. Unlike CliZ that pursues a good balance between compression speed/throughput and quality, Huang et al.'s work \cite{hh-nn} focuses on maximizing the compression ratio and quality for climate data. Specifically, they propose a coordinate-based neural network which is trained to overfit the data also combined with the interpolation and quantization techniques, the resulting weight parameters serve as a compressed data for the original climate/weather grid dataset. The proposed method achieves exceptionally high compression ratios, reaching up to 3000$\times$, significantly outperforming SZ3’s compression ratio of 300$\times$, while also yielding lower mean squared error (MSE). However, this improvement comes at the cost of substantially higher computational demands. Specifically, training the network for a single climate dataset required approximately 8 hours using 4 NVIDIA RTX 3090 GPUs, whereas SZ3 completed the compression in just a few minutes using 32 CPU cores.


\vspace{2mm}
\noindent\textbf{6.5 Compression for Cosmology Research}

Modern cosmological simulations are used by researchers and scientists to investigate new fundamental astrophysics ideas, develop and evaluate new cosmological probes, assist in large-scale cosmological surveys, and investigate systematic uncertainties~\cite{heitmann2019hacc, friesen2016situ}.
Historically such studies have required large computation- and storage-intensive simulations that are  run on leadership supercomputers.
In order to adapt to this evolution, cosmological simulation codes such as Nyx~\cite{almgren2013nyx} (an adaptive mesh cosmological simulation code) have been designed to take advantage of GPU-based HPC systems and can be efficiently scaled to simulate trillions of particles on millions of cores~\cite{almgren2013nyx}. These simulations often periodically dump raw simulation data to the storage for future post hoc analysis.
With the increase in scale of such simulations, saving all the raw data generated to disk becomes impractical because of limited storage capacity and  bottlenecks in the simulation due to the I/O bandwidth required to save the data to disk  ~\cite{wan2017comprehensive,wan2017analysis,cappello2019use}.

Research has shown that general-purpose data distortion metrics, such as \textit{peak signal-to-noise ratio} (PSNR), \textit{normalized root-mean-square error}, \textit{mean relative error}, and \textit{mean squared error}, on their own cannot satisfy the demand of quality for cosmological simulation post hoc analysis~\cite{jin2020understanding,grosset2020foresight}.
Additionally, approaches  utilizing lossy compression for scientific datasets usually apply the same compression configuration to the entire dataset~\cite{jin2020understanding, tao2017exploration}. Yet not all partitions (regions) in the cosmological simulation have the same amount of information.
Cosmologists are typically interested in the dense regions since these  contain halos (clusters of particles) where galaxies are formed. 

To significantly improve the compression performance and control the compression error for cosmological data, Jin et al.~\cite{jin2021adaptive} introduced an adaptive approach to select feasible error bounds for different partitions, showing the possibility and efficiency of adaptively configuring lossy compression for each partition individually.
Specifically, the authors built analytical models to estimate the overall loss of post-analysis results due to lossy compression and to estimate compression ratio, based on the property of each partition.
Then, they used an efficient optimization method to determine the best-fit configuration of error bounds combination in order to maximize the compression ratio under acceptable post-analysis quality loss.
The work introduces negligible overheads for feature extraction and error-bound optimization for each partition.
Overall, this fine-grained adaptive configuration approach improves the compression ratio by up to 73\% with the same post-analysis distortion with only 1\% performance overhead.

More recently, Jin et al.~\cite{jin2022accelerating, jin2024concealing} proposed that the parallel write performance of cosmological data can be significantly improved by a parallel write solution that deeply integrates predictive lossy compression with the asynchronous I/O feature in HDF5. It uses a more advanced ratio-quality model to accurately predict the compression ratio of all partitions and estimate the offsets to allow overlapping between compression and I/O. 
Evaluation shows that, with up to 4,096 cores from Summit, this solution improves the write performance by up to 4.5× and 2.9× over the non-compression and lossy compression filter solutions, respectively, with only 1.5\% storage overhead (compared with original data) on cosmological simulation.

\vspace{2mm}
\noindent\textbf{6.6 Compression for Topological Data Analysis}

Topological data analysis is essential in abstracting, summarizing, and understanding scientific data in various applications, ranging from cosmology and combustion to Earth simulations and AI.  Topological feature descriptors, or simply topological descriptors, provide robust capabilities for capturing, summarizing, and comparing features in scientific data.  Most lossy compressors cannot preserve topological features, thus not guaranteeing topology preservation in decompressed data.  
Below is a review of customized lossy compressors in terms of topological data analysis from two aspects: scalar field topology and vector field topology.

Scalar field topological descriptors include persistence diagrams, merge trees, contour trees, Reeb graphs, and Morse and Morse--Smale complexes.   Key constituents of these descriptors include critical points (maxima, minima, and saddles) and their relationships.  Earlier in 2018, Soler et al.~\cite{SolerPCT18} developed a method to adaptively quantize data based on a given persistent simplification threshold $\epsilon$.  This method guarantees the preservation of critical point pairs with a persistence larger than $\epsilon$ yet does not enforce pointwise error control if the persistence threshold is larger than user-specified error bounds.  More recently, Yan et al.~\cite{topoSZ} proposed TopoSZ, which builds on top of SZ1.4 with a customized quantization scheme to allow different lower/upper bounds per point based on the segmentation induced by contour trees.  TopoSZ also iteratively tests whether there are false-positive/false-negative critical points in decompressed data until convergence. 

Vector fields are a common output form in scientific simulations, such as fluid dynamics, climate and weather, and tokamak simulations.  Topological features of vector fields, such as critical points, separatrices, and critical point trajectories, are crucial to structural understanding and  thus must be preserved in vector field compression.  Until recently, little research had been done  on the preservation of vector field features. In 2020 and later, however,  Liang et al.~\cite{LiangGDCRLOCP20, LiangDCRLOCPG23} proposed cpSZ to preserve all critical points in a vector field without false-negatives, false-positives, and false-types.  A false-negative means the critical point appeared in the original data but was missed in the decompressed data in the exact cell location; a false-positive means an artificial critical point is introduced in the decompressed data but does not exist in the original data; a false-type indicates that although the same critical point exists in both original and decompressed data, the type of the critical point (e.g., source, sink, or saddle) is wrong in the decompressed data.  Specifically, cpSZ derives an analytically sufficient error bound for each point such that no false cases exist in the decompressed data.
This approach has been extended to preserve critical points extracted by simulation of simplicity (SoS)~\cite{edelsbrunner1990simulation}, a more robust critical point extraction algorithm than the numerical one. 
Since SoS relies on the signs of determinants to determine the existence of critical points in a cell, the extended version of cpSZ~\cite{xia2024optimizing} establishes the theory for preserving signs of determinants in lossy compression and leverages it to preserve critical points in vector fields. 
To achieve high compression ratios, relaxation strategies on the derived error bound are explored in the sequential algorithm, and a ghost-aware parallelization strategy is proposed for execution on distributed-memory systems. 

\vspace{2mm}
\noindent\textbf{6.7 Compression for Multi-resolution Data}

Multi-resolution methods, such as the Adaptive Mesh Refinement (AMR) technique have been widely used in scientific simulations~\cite{zhang2019amrex,stone2020athena++,Flash-X-SoftwareX}.  AMR aims to reduce computational expenses while preserving the accuracy of simulation outcomes. Unlike traditional uniform mesh techniques that apply consistent resolution throughout the simulation space, AMR employs a dynamic approach. It selectively increases resolution in regions of interest, thereby optimizing computational resource usage and minimizing storage requirements. \textit{However, the space saved from using multi-resolution alone is often not enough.}
For instance, a multi-resolution dataset with $0.5 \times 1024^3$ mesh points at the coarse level and $0.5 \times 2048^3$ at the fine level could yield about 1 TB of data per snapshot.
Consequently, conducting five simulations each with 200 snapshots require a total storage of 1 PB.


Luo~\textit{et~al.} introduced zMesh~\cite{zMesh}, which pre-processes/reorders AMR data as a 1D array across different refinement levels using z-ordering, thereby leveraging data redundancy between these levels. However, by compressing the data in a 1D array, zMesh is unable to exploit higher-dimensional compression, leading to the loss of spatial information in higher-dimensional data.

Wang~\textit{et~al.} propose TAC~\cite{Wang_hpdc2022} to leverage high-dimensional SZ compression for each refinement level of AMR data. Specifically, to handle AMR data's hierarchical and sparse nature, TAC introduces three pre-processing strategies and adaptively applies them based on the data's characteristics. These strategies include: (1) an optimized sparse tensor representation (OpST) for \textit{low-density} AMR levels; (2) an enhanced $k$-d tree approach for \textit{medium-density} AMR levels; and (3) a padding approach (GSP) for \textit{high-density} AMR levels.
TAC was then extended TAC to TAC+~\cite{Wang_tpds2024} by specifically improving SZ2's performance on AMR data through Shared Huffman Encoding (SHE). This approach enables individual predictions for each small block while using a single shared Huffman tree for encoding, which enhances prediction accuracy and CR for the small blocks generated during the pre-processing of AMR data.

While zMesh and TAC(+) offer offline compression solutions for AMR data, they did
not delve into in-situ compression, which could notably reduce the I/O cost. AMReX supports in situ AMR compression, while it compresses high-dimensional data in 1D and utilizes a small HDF5 chunk size, leading to lower CR and reduced I/O performance.
To address these issues, Wang~\textit{et~al.} propose an effective in situ lossy compression framework for AMR simulations, AMRIC~\cite{amric}, which enhances I/O performance and compression quality. Unlike AMReX's naïve in situ compression approach, AMRIC can perform 3D compression and significantly improve SZ2's compression performance on AMR data based on TAC+. Additionally, AMRIC incorporates the HDF5 compression filter to further enhance compression, I/O performance, and usability.


\vspace{2mm}
\noindent\textbf{6.8 Compression for Seismic Imaging}

Seismic imaging is a technique for determining the seismic properties of the Earth's subsurface~\cite{li2022research}. The technique is extensively utilized in earthquake imaging and resource exploration, including hydrocarbon and geothermal, by energy companies such as Saudi Aramco~\cite{huang2023towards}.
Reverse time migration (RTM) is a cutting-edge seismic imaging method since it can effectively analyze complex seismic structures (e.g., complex velocity focusing and steep (>70$^{\circ}$) dips imaging), compared with traditional methods such as Kirchhoff and wave equation migration ~\cite{farmer2009ss}.

A notable limitation of RTM is the massive data it generates during its execution.
In general, RTM is a full two-way wave equation and can be explained as follows.
Once the input data and configurations, such as the velocity model, are prepared, RTM conducts a \textbf{forward propagation} using the seismic waves. This phase typically involves thousands of time steps, with each step producing a single snapshot.
After this phase, RTM performs a \textbf{backward propagation} based on the reverse order of the generated snapshots, creating the final stacking image, which represents the overall seismic structure.
In real-world use cases, a 10$\times$10$\times$8 cubic kilometers geological structure may produce up to 2,800 TB of data within only a single time step~\cite{RTM-tutorial}.
Storing such big data into peripheral devices can degrade the runtime performance drastically, which motivates error-bounded lossy compression a promising solution to reduce the memory footprint~\cite{huang2023towards,barbosa2023reverse}.

Huang et al.~\cite{huang2023towards} proposed a hybrid lossy compression method called HyZ that combines blockwise regression (BR) and SZx~\cite{szx} to improve the performance of RTM overall execution. Evaluation on 3,600 snapshots of the Overthrust model shows HyZ achieves a compression ratio of 12.31x and compression/decompression speeds of 10.69 GB/s and 12.45 GB/s, respectively. Integration of HyZ into an industrial parallel RTM code improves overall performance by 6.29--6.60x over the execution without compression techniques, outperforming second-tier compressors like SZ and ZFP by up to 2.23x. HyZ also demonstrates higher fidelity than BR in preserving the visualization quality of single snapshots and the final stacking image.

Barbosa et al.~\cite{barbosa2023reverse} introduced an on-the-fly lossy and lossless wavefield compression strategy for RTM to reduce the computational cost and storage demand. They leveraged the ZFP and Nyquist sampling theorem to compress the source wavefield solution before storage and decompress it during the imaging condition calculation. Experimental results on 2D and 3D benchmarks show the seismic image quality is preserved with compression ratios up to 18.84x and 2.08x, respectively. Computational tests using 24 CPU cores and 4 GPUs indicate that the overhead of compression ranges from 122\% to 381\% of the baseline RTM runtime but allows reducing storage by up to 66.7\%. The proposed integration of wavefield compression in RTM enables substantial reductions in I/O and storage needs with minimal impact on image accuracy.

\vspace{2mm}
\noindent\textbf{6.9 Compression for X-ray Light Source Data}

Light sources such as the Advanced Photon Source (APS) at Argonne National Laboratory and the Linac Coherent Light Source (LCLS) at SLAC National Accelerator Center produce enormous volumes of data.  With the completion of the APS upgrade project and the LCLS 2 high energy projects these systems are expected to produce data at rates exceeding 1 TB/s for some experiments and beamlines.  This deluge of data presents a monumental challenge to move the data within and between sites and store the data for archival purposes.
In many cases, a compression ratio target of a $10\times$ is desired for online workflows \cite{roibin}.

So far, compression for light sources has been extensively studied in the fields of ptychography and serial crystallography.
As with other disciplines, the quality  of the decompressed data is of the upmost importance.
However, a key challenge in assessing the quality is the automation of the analysis techniques used to study light source data---in many of these domains, the evaluation of datasets is still largely a time-consuming, manual process \cite{roibin}.  More work to automate these workflows would accelerate the development of compressors for these applications.

For ptychography, the current state of the art is expressed in \cite{sz3}.
These data present as 2D float-encoded integer data recorded over time for a third dimension.\footnote{The detectors used in beamlines often produce unsigned 14- or 16-byte integer data; but after gain correction, pedestal correction, and calibration the data take the form of single-precision floating-point data.  An open research question ia whether compression can be effectively perform on raw data without these steps.}
In this work, a pipeline is constructed in SZ3 that uses different prediction schemes based on the error bound.
At higher error bounds, a multidimensional regression predictor is used.
At lower error bounds, a specialized 1D Lorenzo predictor is used on a transposed version of the 3D input data that aligns all time steps of a particular pixel consequently in memory.
The 1D Lorenzo prediction results in higher quality because spatially adjacent pixels may or may not actually be correlated, resulting in lower quality when using them for prediction.
Together, this pipeline achieves higher rate-distortion results than any other variant of SZ, which was the prior state of the art of  these data.
While these results present high quality at each bit rate, they were  evaluated  by using only traditional rate-distortion curve measures, leaving room for evaluations using more domain-specific metrics.

For serial crystallography, two major approaches can be combined: non-hit rejection and ROIBIN-SZ \cite{roibin}.
Like ptychography, the data present as 2D float-encoded integer data over a time dimension.
Unlike ptychography, however, the data are substantially  noisy, and there are features called Bragg spots or peaks that are key to the posthoc analysis.
In order to achieve high compression ratios, a method called ROBIN-SZ was developed.
ROIBIN-SZ uses the peak information used to perform non-hit rejection to losslessly preserve rectangular regions around the Bragg peaks, while using aggressive $2 \times 2$ binning followed by SZ3 compression on the background.
Preserving the background is critical because the peak-finding process is not infallible.  There could be false-negative peaks to preserve in the dataset in order for the analysis process to complete.

\vspace{2mm}
\noindent\textbf{6.10 Compression for Data Transfer over WAN}

Recently,  error-bounded lossy compression techniques have also been used to improve the data transfer performance over the wide area network (WAN). Ocelet \cite{ocelet}---a lossy-compression-based data transfer accelerator developed for the Globus platform---is a typical example. The Ocelet framework is composed of 8 components/modules:  user interface, FuncX service \cite{funcx}, Globus service \cite{globus}, parallel executor, MPI  call module, error-bounded lossy compression module, data loader/writer, and lossy compression quality estimation module. 
The parallel executor is used to launch the compression/decompression work in a parallel job. Globus manages the data transfer. FuncX service deals with remote orchestration. The user interface offers a graphical interface that helps users submit the tasks easily.

In addition, the Ocelet framework employs optimization strategies to enhance data transfer performance, addressing I/O contention, compute-node delays, and small-file transfer slowdowns. To avoid overloading data transfer or login nodes, Ocelet uses a sentinel program to dynamically monitor and schedule compression tasks adaptively. Initial data transfers occur immediately upon request, while remaining data is compressed and transferred once compute resources are allocated by the scheduler. This adaptive parallel compression significantly improves performance, reducing transfer times by over 90\% in Globus WAN experiments \cite{ocelet}.



\vspace{2mm}
\noindent\textbf{6.11 Compression for Boosting Communication in HPC Clusters}

Researchers have been actively investigating the application of lossy compression to boost the performance of communication in high-performance clusters, focusing on two primary categories: point-to-point and collective communication. 

In point-to-point communication, Zhou et al.~\cite{Zhou2021GPUCOMPRESSION}  utilized 1D fixed-rate ZFP compression \cite{cuZFP} to enhance MPI communications within GPU clusters. This method predominantly enhances point-to-point communication effectiveness but falls short in collective communication scenarios. Moreover, its fixed-rate design, which favors compressed data size over accuracy, fails to assure bounded error, a crucial aspect in lossy compression.

On the collective communication front, Huang et al. \cite{huang2023ccoll,hzccl} developed CCOLL and HZCCL framework to improve performance across all MPI collectives. Such CPU-based methods are a significant advancement in compression-enabled MPI collective communication, demonstrating significant performance improvement over traditional MPI collectives and various baselines. They also provided both theoretical analysis and experimental results to prove the limited impact of error-bounded lossy compression on the final accuracy of collective communications. 

Addressing collective communication on GPU clusters, Zhou et al.~\cite{Zhou2022GPUCOMPRESSIONALLTOALL} enhanced MPI\_Alltoall performance on GPUs through 1D fixed-rate ZFP. Their method, however, depends on a CPU-centric staging algorithm tailored for a singular collective operation, thus limiting its applicability and performance. The fixed-rate compression further compounds these limitations, affecting both performance and compression quality.
In response, Huang et al.~\cite{huang2023gzccl, huang2024optimizing} presented a GPU-centric framework designed to optimize both collective computation and data movement, while efficiently controlling data distortion. This innovative approach harnesses the full computational capabilities of GPUs, significantly reducing the compression cost, synchronization, and device-host data transfers. The resulting performance improvements are notable, surpassing NCCL and Cray MPI by up to 4.5$\times$ and 28.7$\times$, respectively.

\vspace{2mm}
\noindent\textbf{6.12 Compression for Distributed Machine Learning \& Federated Learning Systems}


Recent years have witnessed the rapid evolution of deep learning models for getting high model accuracy, especially in the realm of large-scale models~\cite{bommasani2022opportunities}. Typical examples include large language foundation models (e.g., Palm and GPT-4~\cite{anil2023palm,openai2023gpt4}), large-scale models in computer vision (e.g., VGGs, ResNets~\cite{simonyan2015deep, he2016deep}), and life science (e.g., AlphaFolds~\cite{alphafold}). 
Distributed systems, including the public/private cloud and HPC, provide strong support for training large-scale models based on large-volume training datasets to accelerate the training procedure ~\cite{zhang2022momentum,zhang2021sapus, xu2024fedfa}. 

Communication between computing nodes during training has been recognized as the primary bottleneck in distributed training systems. During the training, gradients, model parameters, and activation data are transmitted across various nodes. As model sizes increase, the volume of this data escalates significantly, resulting in substantial communication overhead. To enhance training performance, it is essential to compress these data to reduce communication demands. 





\textbf{\textit{Gradient Sparsification}} The core idea here is transmitting only the gradients, which play a significant role in the model update \cite{zhang2022mipd}, such as Top-K sparsification \cite{aji2017sparse, lin2017deep, renggli2019sparcml}. \textbf{Gradient Quantization} These approaches use the low-precision data to represent the original data (e.g., defined by float32 data type), which maps the discretized continuous value to different integers in a range, such as one-bit SGD \cite{seide20141}, signSGD \cite{bernstein2018signsgd}, TerGrad \cite{wen2017terngrad}, and QSGD \cite{alistarh2017qsgd}. \textbf{Low Rank} Recent works find that the learned model has a ``low stable rank'' for the modern overparameterized DNN models \cite{martin2021implicit, li2018algorithmic, vogels2019powersgd}, which can explain the impressive generalization properties of the trained DNN models. This opens an opportunity for gradient compression using the low-rank approaches.




\textbf{\textit{Error-Bounded Lossy Compressors}} An emerging area of research is using error-bounded lossy compressors to reduce the size of model updates for distributed systems. Early approaches such as DeepSZ~\cite{DeepSZ} focused on providing a framework for compression and storage of general DNN model architectures. This work showed that through compressing the weights of the model, high compression ratios could be achieved, leading to reported $\approx 50\times$ compression on the weights. Other studies, such as the FedSZ framework~\cite{wilkins2023efficient}, demonstrate how SZ-based compression can be integrated into federated learning (FL). By compressing local model updates before transmission, FedSZ effectively reduces the bandwidth requirements and latency in FL systems, particularly in edge computing scenarios. A significant advantage of using error-bounded lossy compression is that it can retain more information than the above methods. While compressors introduce noise, they do not erase information like sparsification or quantization at certain error boundaries. Developing error-bounded lossy compression strategies to target communication reduction for distributed learning systems is an open area of research. 

\vspace{2mm}
\noindent\textbf{6.13 Compression for Other Applications} 

In addition to the above-listed 12 representative scientific applications, there are more scientific applications that may generate masses of data to store and analyze potentially, including Plasma Physics Simulations (PPS) and  Computational Fluid Dynamics (CFD). PPS, especially those employing Particle-in-Cell (PIC) methods commonly used in fusion energy research, can generate massive amounts of data (from tens to hundreds of terabytes per simulation). This is primarily due to the requirement to accurately model kinetic behaviors by tracking billions or even trillions of particle states across high-dimensional phase spaces (including positions, velocities, and electromagnetic fields). Recent research has shown that advanced compression approaches, including neural-network-based techniques (e.g., variational autoencoders) \cite{choi2021neural} and unsupervised machine-learning checkpoint strategies utilizing Gaussian Mixture Models \cite{CHEN2021110185}, hold great promise for alleviating storage and I/O challenges, facilitating efficient data handling, and enabling effective scientific analysis at extreme computing scales. As for CFD, it may also produce vast volumes of data to store and analyze, in that it typically requires very fine computational grids to accurately capture complex fluid phenomena (e.g., turbulence, shock waves, boundary layers) and involves quite a few physical variables (velocity components, pressure, density, temperature, turbulence quantities) at every grid point. Some compression methods have been developed for the compression of CFD data. BigWhoop \cite{bigwhoop}, for instance, is a compression library for numerical datasets. Lee et al. \cite{lee2024machinelearningtechniquesdata} also explored machine Learning techniques for data reduction of CFD applications.
\section{Conclusion and Future Work}
\label{sec:conclusion}

In this paper we provide a comprehensive survey to discuss lossy compressors for scientific datasets in multiple facets. The key contributions include a novel compression model taxonomy, comprehensive discussion of modular compression techniques and analysis of both general-purpose scientific lossy compressors and domain-specific customized compressors.  

The key takeaways from this survey include: (1) Most state-of-the-art lossy compressors have distinct strengths and weaknesses, as highlighted in our model taxonomy (Figure \ref{fig:comp-taxo}) and the overview of general-purpose compressors (Table \ref{tab:pipelines}), particularly in terms of throughput, compression ratios, and quality. (2) Modern lossy compressors typically rely on a combination of multiple compression techniques to achieve optimal performance, as demonstrated in Table \ref{tab:pipelines}, so deeply understanding each compression technique is critical to the development of effective compression algorithm. (3) Customizing lossy compressors for specific applications or use cases presents a promising avenue for development, as this can leverage the unique characteristics of specific datasets and meet user requirements can lead to highly effective compression solutions. (4) Selecting optimal compression models and approaches for diverse datasets, while tuning parameters like error bounds to meet complex user requirements for data quality, remains a challenging and active research area.

In the future, we aim to expand our survey of modern error-bounded lossy compression techniques, with a particular focus on their application across diverse devices, including emerging accelerators such as FPGAs, GPUs, and CPU vectors. Another promising avenue for exploration is the development and utilization of lossless compression techniques in scientific applications. This would complement our current survey, which primarily emphasizes lossy compression methods. In addition, most of the existing lossy compressors or compression techniques (such as prediction and wavelet transform) are depending on high autocorrelation of the data, so how to develop effective compressors for the datasets with low autocorrelations could be a future research direction.

\section*{Acknowledgments}

This research was supported by the U.S. Department of Energy, Office of Science, Advanced Scientific Computing Research (ASCR), under contract DE-AC02-06CH11357 and DE-SC0024559, and supported by the National Science Foundation under Grant OAC-2104023, OAC-2311875, OAC-2311876, OAC-2311878, OAC-2330367, OAC-2311756, OAC-2311878, OAC-2313123, OAC-2344717,  SHF-1943114, SHF-1910197, OAC-2313122 and NSF-2211538.  The authors extend their appreciation to the Deanship of Graduate Studies and Scientific Research at University of Bisha for funding this research through the promising program under grant number (UB-Promising -  40 -1445).


\end{document}